  \documentclass[a4paper]{article}
  \usepackage[affil-it]{authblk}
  \usepackage{etoolbox}
  \newbool{PREPRINT}
  \booltrue{PREPRINT}
\usepackage[cm]{fullpage}
\usepackage[colorlinks,bookmarksopen,bookmarksnumbered,citecolor=red,urlcolor=red]{hyperref}
\usepackage{amssymb,amsmath}
\usepackage{stmaryrd}
\usepackage{graphicx}
\usepackage{algorithm,algorithmic}

\renewcommand{\vec}[1]{\ensuremath{\boldsymbol{#1}}}
\newcommand{\mat}[1]{\ensuremath{\underline{#1}}}

\newcommand{\rhat}{\ensuremath{\hat{\vec{r}}}}

\newcommand{\bnabla}{\boldsymbol{\nabla}}

\newcommand{\order}{\mathcal{O}}

\newcommand{\Rearth}{R_{\operatorname{earth}}}

\newcommand{\figdir}{./}

\newcommand{\Surface}{\mathcal{S}}
\newcommand{\hS}{{\Surface}}
\newcommand{\nablaS}{\ensuremath{\bnabla_{\!\!\Surface}}}

\newcommand{\rhorelax}{\rho_{\operatorname{relax}}}
\newcommand{\kSmooth}{\ensuremath{\operatorname{Smooth}}}
\newcommand{\kResSmooth}{\ensuremath{\operatorname{RestrictSmooth}}}
\newcommand{\kRes}{\ensuremath{\operatorname{Residual}}}
\newcommand{\kResNorm}{\ensuremath{\operatorname{\text{Residual norm}}}}
\newcommand{\kProlongate}{\ensuremath{\operatorname{Prolongate}}}
\newcommand{\kSpMV}{\ensuremath{\text{(Fused) SpMV}}}
\newcommand{\kTridiag}{\ensuremath{\text{(Fused) Tridiag}}}
\newcommand{\FLOPs}{\ensuremath{\operatorname{FLOPs}}}

\newcommand{\Mem}{\ensuremath{\operatorname{Mem}}}
\newcommand{\MemC}{\ensuremath{\operatorname{Mem}^{(C)}}}

\newcommand{\halosize}{\ensuremath{\text{\texttt{halosize}}}}
\newcommand{\OL}{\ensuremath{\text{\texttt{OL}}}}
\newcommand{\BWmem}{BW_{\operatorname{mem}}}
\newcommand{\BWMPI}{BW_{\operatorname{MPI}}}

\newcommand{\PFLOPs}{\ensuremath{\operatorname{PFLOPs}}}
\newcommand{\TFLOPs}{\ensuremath{\operatorname{TFLOPs}}}
\newcommand{\GBytes}{\ensuremath{\operatorname{GByte/s}}}
\newcommand{\linIdx}{\ensuremath{\Lambda}}
\newcommand{\gridHoriz}{\ensuremath{\mathcal{T}_h}}
\newcommand{\Idnz}{\ensuremath{\mathbb{I}_{n_z\times n_z}}}

\begin{document}
\title{Petascale elliptic solvers for anisotropic PDEs on GPU clusters}

\ifbool{PREPRINT}{ 
\author[1,*]{Eike Hermann M\"{u}ller}
\author[1]{Robert Scheichl}
\author[2]{Eero Vainikko}
\affil[1]{Department of Mathematical Sciences, University of Bath, Bath BA2 7AY, United Kingdom}
\affil[2]{Institute of Computer Science, University of Tartu, Liivi 2, Tartu 50409, Estonia}
\affil[*]{Email: \texttt{e.mueller@bath.ac.uk}}
\maketitle %
}{
\begin{frontmatter}
\author[addressBath]{Eike Hermann M\"{u}ller\corref{corauthor}}
\ead{e.mueller@bath.ac.uk}
\cortext[corauthor]{Corresponding author}
\author[addressBath]{Robert Scheichl}
\author[addressTartu]{Eero Vainikko}
\address[addressBath]{Department of Mathematical Sciences, University of Bath, BA2 7AY, Bath, United Kingdom}
\address[addressTartu]{Institute of Computer Science, University of Tartu, Liivi 2, Tartu 50409, Estonia}
} 

\begin{abstract}
Memory bound applications such as solvers for large sparse systems of equations remain a challenge for GPUs. Fast solvers should be based on numerically efficient algorithms and implemented such that global memory access is minimised. To solve systems with up to one trillion ($10^{12}$) unknowns the code has to make efficient use of several million individual processor cores on large GPU clusters.

We describe the multi-GPU implementation of two algorithmically optimal iterative solvers for anisotropic elliptic PDEs which are encountered in atmospheric modelling. In this application the condition number is large but independent of the grid resolution and both methods are asymptotically optimal, albeit with different absolute performance. We parallelise the solvers and adapt them to the specific features of GPU architectures, paying particular attention to efficient global memory access. We achieve a performance of up to 0.78 PFLOPs when solving an equation with $0.55\cdot 10^{12}$ unknowns on 16384 GPUs; this corresponds to about $3\%$ of the theoretical peak performance of the machine and we use more than $40\%$ of the peak memory bandwidth with a Conjugate Gradient (CG) solver. Although the other solver, a geometric multigrid algorithm, has a slightly worse performance in terms of FLOPs per second, overall it is faster as it needs less iterations to converge; the multigrid algorithm can solve a linear PDE with half a trillion unknowns in about one second.
\end{abstract}
\ifbool{PREPRINT}{}{%
\end{frontmatter}
} 
\section{Introduction}
Many problems in geophysical modelling require the fast solution of anisotropic partial differential equations (PDEs) in ``flat'' domains. For example, a global PDE for the pressure correction has to be solved in every time step of many numerical weather- and climate prediction models if an implicit method is used to advance the atmospheric fields forward in time. As the height of the atmosphere is much smaller than the horizontal extent of the domain in global models, after discretisation this equation has a very strong vertical anisotropy. The discretised PDE can be written as a sparse system of equations
\begin{equation}
  \mat{A}\vec{u} = \vec{f}, \label{eqn:SparseSystemSchematic}
\end{equation}
where $\mat{A}$ is a sparse $n\times n$ matrix and the vector $\vec{u}$ has $n$ entries and represents the global (Exner-) pressure correction field in the whole domain. In modern applications the number of degrees of freedom $n$ can be very large, for a global grid with a horizontal resolution of $1km$ or less and $\order(100-200)$ vertical levels a system with $n\gtrsim 10^{11}$ unknowns needs to be solved. Resolutions of this order are expected to be achieved by state-of-the-art global forecast models within the next decade. As the PDE is solved in every model time step and accounts for a significant amount of the total model runtime, it is crucial to solve it as fast as possible to deliver forecasts on operational timescales. This can only be achieved by using algorithmically optimal methods and implementing them on the fastest available hardware. The main challenge faced by the solver is the vertical anisotropy which prevents a standard approach such as a geometric multigrid algorithm with point smoother or an iterative solver with Jacobi preconditioner. Recently we have shown that Krylov subspace solvers and multigrid methods tailored to the structure of the problem are highly efficient and scale up to tens of thousands of CPU cores \cite{Mueller2013b}.
In particular we found that for our application the tensor-product geometric multigrid solver, suggested and first analysed in \cite{BoermHiptmair1999}, is superior to a preconditioned Conjugate Gradient method and to state of the art parallel algebraic multigrid (AMG) implementations from the DUNE \cite{Blatt07,Blatt10} and Hypre \cite{Henson2000} libraries.

As explained in detail below, the elliptic PDE for the atmospheric pressure correction has the structure of a shifted Laplace equation, which is usually known as the (sign-positive) Helmholtz equation in the meteorological literature. In contrast to the conventional Helmholtz equation encountered for example in wave scattering problems, the elliptic operator we consider is positive definite. While implicit time marching schemes permit larger model time steps, advective time scales and constraints on the accuracy of the solution limit the permitted time step size. This implies that the CFL number, which is proportional to the ratio of the time step size and horizontal grid spacing, is typically in the range $2-10$. After preconditioning, the condition number of the elliptic operator is $\order(100-1000)$, independent of the horizontal grid resolution. Because of this Krylov subspace methods and multigrid algorithms with a fixed number of levels are asymptoptically optimal and algorithmically scalable as the problem size increases.

Graphics Processing Units (GPUs) have been used very successfully in many areas of Scientific Computing and can be superior to more traditional CPU architectures both in terms of speed and power efficiency. A particular challenge for solvers of sparse systems of linear equations such as (\ref{eqn:SparseSystemSchematic}) is that their performance is typically limited by the speed with which data can be read from (and written to) global GPU memory. While the number of floating point operations for the iterative solvers we consider is typically two to five times larger than the number of memory operations, on modern GPUs, such as the Kepler GK110 on the K20X cards on the Titan supercomputer \cite{top500.org} the cost for one (double precision) memory access is more than $40\times$ larger than the cost of one floating point operation. This factor is given by the ratio of the peak floating point performance and the peak global memory bandwidth, namely $1.31 \TFLOPs/(250\GBytes) \times 8 \operatorname{Byte} \approx 42$ for double precision arithmetic on the K20X card \cite{TeslaDatasheet}. It should be compared to the corresponding number for CPU architectures where only around 3 floating point operations can be carried out per variable loaded from memory. Furthermore, due to limited memory, only problems with up to a few million degrees of freedom can be solved on a single GPU. To solve larger systems a distributed-memory multi-GPU implementation has to be used. For problems with up to a trillion ($10^{12}$) unknowns, several million processor cores are necessary.

To implement the fastest possible massively parallel GPU solver we followed three design principles:
\begin{enumerate}
\item \textbf{Algorithmically optimal solver.}
To minimise the overall solution time, the biggest gains can be achieved by using an iterative solver method which is tailored to the problem to be solved and converges in the smallest possible number of iterations. 
Krylov subspace methods are very popular in meteorological applications because of their simplicity (see e.g. \cite{Skamarock1997,Thomas1997,Qaddouri2003,Davies05} and the detailed review in \cite{Mueller2013b}). 
For anisotropic PDEs it is particularly important to exploit the strong coupling in the vertical direction by using a suitable preconditioner. 
Since the elliptic system considered in this work is symmetric and positive definite, the most suitable Krylov subspace method is a Conjugate Gradient solver preconditioned with vertical line relaxation. The preconditioner requires the frequent solution of a tridiagonal system in each vertical column; this can be achieved with the Thomas algorithm (see e.g. \cite{Press2007}).

However, we already found in \cite{Mueller2013b} that the geometric tensor-product multigrid solver proposed and analysed in \cite{BoermHiptmair1999}, which uses vertical line relaxation as the smoother, converges significantly faster than the preconditioned CG iteration. As the numerical experiments in this article confirm, the CG solver requires at the order of 60 iterations to reduce the residual by five orders of magnitude, whereas the multigrid method converges in less than 10 iterations.

\item \textbf{Memory optimised CUDA-C implementation.}
We optimised the single GPU implementation by minimising the number of memory references per iteration.
As already shown in \cite{Mueller2013a}, the biggest gains can be achieved by using a ``matrix-free'' approach and recomputing the (sparse) matrix $\mat{A}$ instead of storing it explicitly. For example, carrying out a sparse matrix-vector product (SpMV) $\vec{y} \mapsfrom \mat{A}\vec{x}$ requires 1 to 7 global reads (depending on the caching of the vector $\vec{x}$) and 1 global write at each grid cell. This should be compared to a matrix-explicit implementation which requires 7 additional reads for a finite volume stencil and can hence be more than twice as expensive. Not storing the matrix explicitly also reduces the memory requirements of the solver significantly. This allows the solution of larger problems and better utilisation of the GPU resources.
For optimal global memory troughput on the GPU it is crucial to adapt the data layout to achieve optimally coalesced access for all threads in a warp. This requires a horizontally contiguous ordering of the degrees of freedom, which differs from the vertically contiguous ordering which allows optimal cache reusage on CPUs. In addition we reduced the number of memory references by fusing several GPU kernels. 
 
We find that on a single GPU our CG implementation achieves $36\%-56\%$ of the peak global memory bandwidth depending on the problem size. For the multigrid solver the rate is slightly lower with $15\%-36\%$ of the peak global memory bandwidth, but this is more than compensated by the faster convergence rate.
\item \textbf{Massively parallel multi-GPU code.}
We extended our implementation to clusters of GPUs by using a horizontal
decomposition of the computational domain, which is common for applications in
atmospheric modelling. For this we used the Generic Communication Library
\cite{bianco2013interface} which allows the straightforward implementation of halo exchanges on structured two- and three-dimensional grids and supports GPUDirect data transfer between different GPUs.

For the largest problem we studied, the additional overhead from the MPI communications is about $10\%$ for the CG solver and about $40\%$ for the multigrid solver, both solvers show very good weak scaling to up to 16384 GPUs.
\end{enumerate}
\paragraph{Main achievements}
In this paper we describe this approach in detail for the solution of a model equation which captures the main features of the elliptic PDE for the pressure correction in global weather- and forecast models; further details on the model equation and relevant meteorological literature can be found in previous publications \cite{Mueller2013b,Dedner2014}.
In \cite{Mueller2013a} we described the single-GPU implementation of a matrix-free CG solver, here we extend this approach to the geometric-multigrid solver analysed in \cite{BoermHiptmair1999} and extend both solvers to run on clusters of GPUs. We tested the performance of our solvers and ran them on up to 16384 GPUs of the Titan Cray XK7 cluster (OLCF, Oak Ridge National Lab), which contains 18,688 nVidia K20X cards with GK110 Kepler GPUs and is currently ranked as the second fastest computer in the world (\texttt{}{top500.org}, June 2014 \cite{top500.org}).

We are able to solve a problem with half a trillion ($0.55\cdot 10^{12}$) degrees of freedom in about one second with the multigrid solver. The GPU implementation is about a factor four faster on one K20X GPU card on Titan than our optimised Fortran 90 CPU code running on one 16 core AMD Opteron processor of HECToR, the UK's national supercomputing resource.

On Titan we achieve a performance of $0.78$ \PFLOPs\ for the CG solver (and $0.65$ \PFLOPs\ for the multigrid algorithm). It should be stressed that this time includes all components of the solver, such as host-device data transfer and transposition of the fields for horizontally contiguous ordering on the GPU. As the code is bandwidth limited, the absolute performance should be quantified in fractions of the peak global GPU bandwidth. For the CG solver we can achieve a percentage of $32\%$-$42\%$ of the peak global memory bandwidth when running on 16384 GPUs, for the multigrid solver this fraction is $15\%$-$25\%$.
All source code is freely available for download under the LGPL 3 license.
\paragraph{Previous work}
Early work on the GPU parallelisation of Conjugate Gradient methods and multigrid solvers for sparse linear systems is discussed in \cite{Bolz03,Goodnight2005}. The authors solve the two dimensional shifted Laplace equation $-\Delta u+\sigma u = RHS$ \cite{Bolz03} and the Poisson equation $-\Delta u=0$ \cite{Goodnight2005} arising in implicit time stepping methods for the solution of the Navier Stokes equations. In both cases the code is implemented by using the low level graphics API on GPU hardware which is quite dated now. More recent GPU implementations of (preconditioned) Conjugate Gradient- \cite{Menon2007,cevahir2009fast,Knittel2010,cevahir2010high,Georgescu2010,griebel2010multi,jacobsen2010mpi} and geometric multigrid solvers \cite{feng2010parallel,Geveler2011,jacobsen2011full} are also reported in the literature. To solve problems which arise for example from finite element discretisations on unstructured grids, typically the system matrix is stored explicitly in formats such as compressed sparse row storage (CSR) or in the ELLPACK format (see \cite{Bell2008} for a detailled discussion of sparse matrix storage formats on GPUs) and the authors concentrate on optimising the sparse matrix-vector multiplication. The advantage of this approach is that the solvers can be applied to very general and grid-independent problems such as power grid simulations in \cite{feng2010parallel} or arbitrary sparse matrices from the University of Florida sparse matrix collection \cite{Davis2011} as described in \cite{cevahir2009fast}. However, for very general problems, the construction of a suitable preconditioner is very difficult and convergence is slow. Similarly, while AMG implementations such as those reported in \cite{haase2010parallel,Brannick2013} can be used to solve a very large class of elliptic problems, the requirement of explicit matrix storage makes them more expensive than geometric multigrid for the structured PDEs which we discuss in this article. On the other hand, the only matrix-free implementations we are aware of are \cite{Goodnight2005,Menon2007,Knittel2010}, and in all cases the authors focus on solving the homogeneous and isotropic Poisson equation in a regular two- or three- dimensional domain. Both extremes should be compared to our approach: by exploiting the structure of the problem to construct a suitable preconditioner our solvers can deal with three dimensional anisotropic equations on curved domains but avoid explicit storage of the matrix which has a negative impact on performance for bandwidth limited applications. 

More recently iterative solvers have also been parallelised across multiple GPUs. For example (unpreconditioned) Conjugate Gradient solvers are tested for a range of sparse matrices in \cite{cevahir2009fast,cevahir2010high,Georgescu2010}. Of particular interest for this work are the multigrid solvers discussed in \cite{jacobsen2010mpi,griebel2010multi,jacobsen2011full} for the 3D Poisson equation which arises in implicit time stepping methods for the solution of the Navier Stokes equations. The equation we consider in the following can be derived in a similar fashion for the compressible Euler equations. A significant but important difference is that compressibility gives rise to an additional zero order term which introduces an intrinsic length scale beyond which interactions between gridpoints are exponentially suppressed. This has an important impact on the parallelisation of the multigrid solver: it is sufficient to use a relatively small number of multigrid levels and one or two smoother iterations are sufficient to solve the well conditioned coarse grid problem, thus avoiding an exact global coarse grid solve across all processors. This should be compared to the more complicated approaches such as parallel coarse grid aggregation discussed in \cite{jacobsen2011full} for the homogeneous Poisson equation.

An interesting approach combining the computational power of both the CPU and the GPU on a node is described in \cite{Goeddeke2008} where the authors describe a BiCG solver for the Poisson problem on an unstructured grid and parallelise the solver on a cluster with up to 17 nodes. A multigrid V-cycle is used for preconditioning and the smoother, which is a separate multigrid iteration on structured subgrids, is offloaded to the GPU. As CPU-GPU data transfer is expensive and there are now efficient ways for exchanging data directly between GPUs on different nodes, we implemented our solver such that all calculations are carried out on the device only.

Only recently clusters with several thousands of GPUs have become available and as far as we are aware to date there are no multi-GPU implementations which have been shown to scale to up to more than around 100 GPUs.
Parallel scaling on up to 128 GPU for a CG solver with Jacobi preconditioner for the Poisson equation is described in \cite{jacobsen2010mpi}, and scaling for the multigrid solver of the same problem is reported on up to 64 nodes in \cite{jacobsen2011full}. Elliptic problems solved so far typically have less than 1 billion unknowns and the results in this work represent a significant contribution to extending the scalability of iterative solvers to several millions of processor cores on tens of thousands of GPUs and for solving very large systems with up to half a trillion unknowns. In this context we mention the work reported in \cite{Yang2013} where the authors achieved a GPU performance of 1.9 \PFLOPs\ for an explicit time stepping solver of the shallow water equations on a cubed sphere grid (however, in contrast to implicit timestepping methods, this does not require the solution of an elliptic PDE for the pressure correction). That solver is run on 3750 nodes with 1 GPU each to solve problems with up to 4 billion unknowns per atmospheric variable.

Although we believe that here we describe the first massively parallel GPU implementation of solvers for sparse systems with more than half a trillion ($0.5\cdot 10^{12}$) unknowns, problems of this size have been solved on more conventional CPU clusters before. For example, a massively parallel implementation of a multigrid solver on hybrid grids is described in \cite{Gmeiner2014} and the authors demonstrate the excellent scalability of the algorithm on nearly 300,000 CPU cores by solving systems with up to $10^{12}$ unknowns.

While in the past, bespoke geometric multigrid solvers for anisotropic elliptic PDEs have been studied extensively in the literature (see e.g. \cite{Trottenberg2001} for a standard textbook), we will not discuss those more algorithmic aspects here and instead refer the reader to to \cite{BoermHiptmair1999} and a forthcoming publication (\cite{Dedner2014}) which contain more compresensive reviews of this topic.
\paragraph{Structure}
This paper is organised as follows:
in \mbox{Section \ref{sec:IterativeSolvers}} we briefly review the application of iterative solvers to anisotropic elliptic PDEs in atmospheric modelling with particular focus on Conjugate Gradient and geometric multigrid methods. The GPU  implementation of these methods is discussed in \mbox{Section \ref{sec:Implementation}} and a theoretical performance analysis is carried out in \mbox{Section \ref{sec:PerformanceAnalysis}}. The results of our numerical experiments and weak scaling tests on Titan are presented in \mbox{Section \ref{sec:Results}}. Finally we conclude and outline some ideas for further work in \mbox{Section \ref{sec:Conclusion}}.
\section{Iterative solvers for anisotropic elliptic PDEs}\label{sec:IterativeSolvers}

\subsection{Model problem}
We consider the following PDE, which can be used as a simplified model of the pressure correction equation arising in semi-implicit semi-Lagrangian time stepping methods in atmospheric forecast models:
\begin{equation}
  - \omega(h)^2 \left(\nablaS^2 u(\rhat,r) +
    \lambda(h)^2\frac{1}{r^2}\frac{\partial}{\partial r}\left(r^2\frac{\partial
    u(\rhat,r)}{\partial r}\right)\right) + u(\rhat,r) = f(\rhat,r).
  \label{eqn:ModelEquation}
\end{equation}
Here $r\in[1,1+H]$ is the radial coordinate in units of the earth's radius $\Rearth$ and $H=D/\Rearth\ll 1$ is the ratio between the depth of the atmosphere and the radius of the earth. The unit vector $\rhat$ is used to describe a position on a unit sphere $\hS$ and $\nablaS \equiv \bnabla - \langle \rhat,\bnabla\rangle\rhat$ denotes the tangential conponent of the three dimensional gradient. Homogeneous Neumann boundary conditions are used at the top and bottom boundary of the domain. A structured vertical grid and a semi-structured horizontal grid $\gridHoriz$ are used for discretising the equation on the domain $\hS\times[1,1+H]$. Since $H\ll 1$ the vertical grid spacing $h_z$ is much smaller than the horizontal mesh width $h$, and so the discretised equation has a very strong grid aligned anisotropy in the vertical direction. The parameters $\omega(h)$ and $\lambda(h)$ depend on the meteorological conditions and on the time step size and are discussed in more details in \cite{Mueller2013b}. Most importantly, as the horizontal resolution increases, i.e. the mesh width $h$ tends to zero, we have $\omega(h)\propto h$ and $\lambda(h)\rightarrow 1$.
More specifically the coefficient of the second order term in (\ref{eqn:ModelEquation}) is given by
\begin{equation}
  \omega(h) = \frac{c_h\Delta t(h)}{2R_{\operatorname{earth}}}
  \label{eqn:omegahdependence}
\end{equation}
where $c_h$ is at the order of the speed of sound. 
Because of fast advective time scales and to represent large scale flow accurately, in meteorological applications the resolution dependent time step size $\Delta t(h)$ has to be chosen such that the horizontal CFL number $\nu_{CFL} = c_h\Delta t/\Delta x = 2\omega(h)/h$ is not larger than around $2-10$. On the other hand, implicit time stepping methods will not be competitive if $\nu_{CFL}$ is too small. 
To satisfy these conditions we always use
\begin{equation}
  \omega(h) = \frac{1}{2}\nu_{CFL}h\qquad\text{with}\quad \nu_{CFL}=8.4
  \label{eqn:OmegaNumerical}
\end{equation}
in our numerical experiments. We also study the robustness of our solvers to variations in the CFL number.

The equation in (\ref{eqn:ModelEquation}) can be seen as a special case of the more general PDE studied in \cite{Dedner2014} (see also \cite{Wood2013} which describes how an equation of this form is derived in the ENDGame dynamical core of the UK Met Office's Unified Model)
\begin{equation}
  -\omega^2 \vec{\nabla} \cdot(\mat{\alpha}(\rhat,r)\vec{\nabla} u(\rhat,r)) - \omega^2 \vec{\xi}(\rhat,r)\cdot \vec{\nabla} u(\rhat,r) + \beta(\rhat,r) u(\rhat,r) = f(\rhat,r),
\label{eqn:ModelEquationComplex}
\end{equation}
where $\mat{\alpha}$, $\vec{\xi}$ and $\beta$ are atmospheric ``profiles'', i.e. functions which depend on the current state of the model. Due to the vertical layering of the atmosphere each of these functions can be approximated very well as the product of a vertically varying field and a horizontally varying function. 
This is why tensor-product methods are of particular interest. Even if the profiles do not factorise exactly, an approximate factorisation can still be used in a preconditioner. This is discussed in a lot of detail in \cite{Dedner2014}. It is for these reasons that we believe that the PDE in (\ref{eqn:ModelEquation}) is a good model for the pressure correction equation encountered in atmospheric models.

We chose not to make any further simplifications such as solving a (shifted) Laplace equation in a simplified geometry, as is often done in the literature on massively parallel solvers, since this would allow significant further performance improvements which are not reasonable in realistic meteorological applications.
\subsubsection{Discretisation}
Equation (\ref{eqn:ModelEquation}) is discretised using a simple cell centred finite volume scheme on one panel of a non-conformal cubed sphere grid with gnomonic projection as described in \cite{Sadourny1972}.

For simplicitly homogeneous Dirichlet boundary conditions are used in the horizontal direction. To represent a field $u(\rhat,r)$, all data in one vertical column above the horizontal grid cell $T$ is stored in a vector $\overline{\vec{u}}^{(T)}$ of length $n_z$. Then the discretised equations associated with the horizontal grid cell $T\in\gridHoriz$ can be written as
\begin{equation}
  \left(\mat{A} \overline{\vec{u}}\right)^{(T)} = \mat{A}_T
  \overline{\vec{u}}^{(T)} + \sum_{T'\in\mathcal{N}(T)} \mat{A}_{TT'}
  \overline{\vec{u}}^{(T')} = \overline{\vec{f}}^{(T)}.
\label{eqn:TridiagonalPDE}
\end{equation}
where $\mat{A}_T$ is a tridiagonal matrix containing all vertical couplings, as well as diagonal terms, and where the diagonal matrices $\mat{A}_{TT'}$ describe the couplings to the horizontally neighbouring cells $T'\in\mathcal{N}(T)$. Due to the strong vertical anisotropy, the entries in $\mat{A}_T$ are much larger than the ones in $\mat{A}_{TT'}$.

To understand the origin of the individual terms in (\ref{eqn:TridiagonalPDE}), it is instructive to write down the explicit form of these matrices for a simplified equation instead of (\ref{eqn:ModelEquation}). Consider the shifted Laplace equation $-\omega^2\left(\boldsymbol{\nabla}_{2D}^2u+\lambda^2\partial^2/\partial_r^2u\right)+u=f$ in a flat box $\Omega \times[0,H]$; here the horizontal domain $\Omega=[0,1]\times[0,1]$ is the unit square and $\boldsymbol{\nabla}_{2D}=\partial^2/\partial_x^2+\partial^2/\partial_y^2$ denotes the two dimensional Laplacian. If we choose an equidistant Cartesian grid with spacing $h$ on $\Omega$ then every horizontal cell $T$ can be labelled with a pair of indices, i.e. $T\equiv (i,j)$. In this case equation (\ref{eqn:TridiagonalPDE}) can be written explicitly as
\begin{equation*}
  \left(\mat{A} \overline{\vec{u}}\right)^{(i,j)} = \mat{A}_{(i,j)}
  \overline{\vec{u}}^{(i,j)}
+ \mat{A}_{(i,j),(i+1,j)}\overline{\vec{u}}^{(i+1,j)} 
+ \mat{A}_{(i,j),(i-1,j)}\overline{\vec{u}}^{(i-1,j)} 
+ \mat{A}_{(i,j),(i,j+1)}\overline{\vec{u}}^{(i,j+1)} 
+ \mat{A}_{(i,j),(i,j-1)}\overline{\vec{u}}^{(i,j-1)} 
= \overline{\vec{f}}^{(i,j)}
\end{equation*}
with the matrices $\mat{A}_{(i,j),(i',j')}=-\omega^2/h^2\Idnz$ where $\Idnz$ is the $n_z\times n_z$ identity matrix. The entries on the diagonal of the matrix $\mat{A}_{(i,j)}$ would be $1+4\omega^2/h^2+2\omega^2\lambda^2/h_z^2$ and the off-diagonal entries are $-\omega^2\lambda^2/h_z^2$ where $h_z$ is the vertical grid spacing. We stress, however, that equation (\ref{eqn:TridiagonalPDE}) allows more general geometries with semi-structured horizontal grids. In this case the exact form of the matrices $\mat{A}_{TT'}$ and $\mat{A}_T$ is more complicated as the finite volume discretisation leads to non-trivial geometric factors.

The elliptic equation in (\ref{eqn:ModelEquation}) is symmetric and positive definite. For a given horizontal resolution we can give a rough estimate of the condition number by again considering the equation in a flat box. After preconditioning by vertical line relaxation, it can be shown that the resolution dependent condition number $\kappa(h)$ is 
\begin{equation}
  \kappa(h) \approx 1 + 8 \omega(h)^2/h^2\qquad\text{for $h\ll 1$ and $\omega(h) \ll 1$}.\label{eqn:ConditionNumber}
\end{equation}
Since $\omega(h)=4.2h$ for $\nu_{CFL}=8.4$, this leads to an estimate of $\kappa(h) \approx 142$. Geometric factors arising from the spherical geometry will modify this estimate by factors of $\order(1)$ and hence we expect the condition number of our problem to be in the range $100-1000$, independent of the grid resolution.

In principle we do not need to make any assumptions on the ordering of the horizontal degrees of freedom and indirect addressing could be used in the horizontal direction, as is described for the DUNE implementation of the problem on quasi-uniform icosahedral and cubed-sphere grids for the entire sphere in \cite{Dedner2014}. However, in this work we assume for simplicity that each horizontal grid cell on the panel can be identified by a pair of indices $(i,j) \in [1,n_x]\times[1,n_y]$, and each vertical level is indexed by an additional integer $k\in[0,n_z-1]$.
\subsection{Algorithmically scalable and efficient solvers}
Large sparse systems of equations can be solved efficiently using state-of-the-art iterative solvers which improve an initial solution $\vec{u}_0$ by reducing the residual $\vec{r}=\vec{f}-\mat{A}\vec{u}$ (and hence the error) at every iteration. Krylov subspace methods (see e.g. \cite{Saad2003} for an introduction) minimise the residual by constructing the solution in the space spanned by the vectors
\begin{equation}
  \vec{r}_0, \mat{A}\vec{r}_0, \mat{A}^2\vec{r}_0, \dots\, ,
\end{equation}
where $\vec{r}_0=\vec{f}-\mat{A}\vec{u}_0$ is the initial residual. The simplest (and most efficient) Krylov subspace method for symmetric positive systems is the preconditioned Conjugate Gradient (CG) iteration. Closely related methods such as Conjugate Residual (CR), GMRES and BiCGStab are very popular in the meteorological literature (see e.g. \cite{Skamarock1997,Thomas1997,Qaddouri2003,Davies05}) and due to the strong vertical anisotropy, a very effective preconditioner $\mat{M}$ is vertical line relaxation, which requires the solution of a tridiagonal problem in each vertical column. This preconditioner corresponds to solving the equation which is obtained by only keeping the first term on the left hand side of (\ref{eqn:TridiagonalPDE}), which describes the dominant vertical couplings; the resulting matrix $\mat{M}$ is block-diagonal.
Each of the tridiagonal systems can be solved independently using the Thomas algorithm (see e.g. \cite{Press2007}). Mathematically this is equivalent to a block-Jacobi or block-SOR method where each of the  blocks correspond to the degrees of freedom in one particular vertical column.

The computationally most expensive components of the algorithm are a sparse matrix-vector (SpMV) multiplication and a preconditioner (tridiagonal-) solve, in the following we write these operations as
\begin{xalignat}{2}
  \vec{y} &\mapsfrom \mat{A}\vec{x}\quad\text{(SpMV)}, &
  \vec{y} &\mapsfrom \mat{M}^{-1}\vec{x}\quad\text{(Preconditioner)}.
  \label{eqn:SpMVPrec}
\end{xalignat}
As discussed in detail in \cite{Mueller2013a} (where also the algorithm is written down explicitly), the efficiency of the implementation can be improved by fusing these two operations with the level 1 BLAS operations in the main loop.
Other Krylov subspace methods, such as BiCGStab, CR or GMRES can be used to solve more general systems and differ from the CG method only in the number of sparse matrix-vector products, preconditioner applications, level 1 BLAS operations, and in the storage requirements.

In contrast, multigrid methods (see e.g. \cite{Briggs2000,Trottenberg2001}) use a hierarchy of coarse levels to minimise the error on all scales simultaneously. In the following we write $\vec{u}^{(\ell)}$ for the field on multigrid level $\ell$, where $\ell=1$ corresponds to the coarsest level and $\ell=L$ to the finest level where we want to solve the equation, i.e. $\vec{u}^{(L)}=\vec{u}$. For simplicity we omit the multigrid-level index $(\ell)$ wherever it is obvious from the context, such as on all the coarse grid matrices. The fine grid equation is solved by starting with an initial guess for the solution and improving on this by repeated calls to the recursive subroutine \texttt{Vcycle} in Algorithm \ref{alg:VCycle} (for simplicity the iteration is written down for one pre- and one post-smoothing step here). After each Vcycle convergence is checked by comparing the norm of the residual to a given tolerance $\varepsilon$, i.e. the algorithm terminates as soon as 
\begin{equation}
||\vec{r}||/||\vec{r}_0||<\varepsilon.\label{eqn:epsilonTolerance}
\end{equation}
In our numerical experiments we always reduce the residual by five orders of magnitude, which is typical in atmospheric applications.
\begin{algorithm}
\caption{Subroutine \texttt{VCycle}($\rhorelax$, $\{\vec{u}^{(\ell)}\}$, $\{\vec{f}^{(\ell)}\}$, $\{\vec{r}^{(\ell)}\}$,$\ell$)}
\begin{center}
 \begin{algorithmic}
 \IF{$\ell = 1$} 
\STATE{\COMMENT{\textit{Restrict residual and solve on coarsest level}}}
   \STATE{$\vec{f}^{(1)}\mapsfrom \mat{R}_{1,2}r^{(2)}$, $\vec{u}^{(1)} \mapsfrom \mat{A}^{-1}\vec{f}^{(1)}$}
 \ELSE
   \IF{$\ell= L$}
		\STATE{}\COMMENT{\textit{Smooth once on finest level}}
      \STATE{$\vec{u}^{(\ell)} \mapsfrom \vec{u}^{(\ell)} + \rhorelax \mat{M}^{-1}(\vec{f}^{(\ell)}-\mat{A}\vec{u}^{(\ell)})$ [= Kernel \kSmooth]}
   \ELSE
     \STATE{}\COMMENT{\textit{On all other levels, restrict residual and smooth once}}
     \STATE{$\vec{f}^{(\ell)}\mapsfrom \mat{R}_{\ell,\ell+1}r^{(\ell+1)}$, $\vec{u}^{(\ell)} \mapsfrom \rhorelax\mat{M}^{-1}\vec{f}^{(\ell)}$ [= Kernel \kResSmooth]}
   \ENDIF
	\STATE{}\COMMENT{\textit{Calculate residual}}
   \STATE{$\vec{r}^{(\ell)}\mapsfrom \vec{f}^{(\ell)}-\mat{A}\vec{u}^{(\ell)}$ [= Kernel \kRes]} 
   \STATE{}\COMMENT{\textit{Recursive call to Subroutine} \texttt{VCycle}}
   \STATE{Call \texttt{VCycle}($\rhorelax$, $\{\vec{u}^{(\ell)}\}$, $\{\vec{f}^{(\ell)}\}$, $\{\vec{r}^{(\ell)}\}$, $\ell-1$)} 
   \STATE{}\COMMENT{\textit{Add prologated coarse grid correction}}
   \STATE{$\vec{u}^{(\ell)} \mapsfrom \vec{u}^{(\ell)}+\mat{P}_{\ell,\ell-1}\vec{u}^{(\ell-1)}$ [= Kernel \kProlongate]}
     \STATE{}\COMMENT{\textit{Postsmoothing}}
     \STATE{$\vec{u}^{(\ell)} \mapsfrom \vec{u}^{(\ell)} + \rhorelax \mat{M}^{-1}(\vec{f}^{(\ell)}-\mat{A}\vec{u}^{(\ell)})$ [= Kernel \kSmooth]}
 \ENDIF
 \end{algorithmic}
  \label{alg:VCycle}
\end{center}
\end{algorithm}
To achieve rapid convergence the different multigrid components have to be adapted to the problem to be solved. In \cite{BoermHiptmair1999} geometric multigrid algorithms for equations with a tensor-product structure and grid-aligned anisotropy are analysed. The authors show that the convergence rate of a multigrid solver for a two-dimensional problem with strong coupling in the vertical direction can be bounded by the convergence rate of the multigrid algorithm for a related one-dimensional horizontal problem if the following tensor-product multigrid algorithm is used:
\begin{itemize}
  \item \textbf{Horizontal-only semicoarsening}: Only coarsen the grid in the horizontal direction.
  \item \textbf{vertical block-Jacobi/-SOR smoother}: Use vertical line relaxation as the smoother, i.e. solve the equation for all degrees of freedom in a vertical column simultaneously. Hence, in essence the multigrid smoother is identical to the preconditioner used for the CG algorithm described above.
\end{itemize}
As shown in \cite{Dedner2014}, the generalisation from two to three dimensions is straightforward and this is the algorithm which we use in this work. On each level we use a block-Jacobi smoother which can be written as
\begin{equation}
  \vec{u}^{(\ell)} \mapsfrom \vec{u}^{(\ell)} + \rhorelax \mat{M}^{-1}\left(\vec{f}^{(\ell)}-\mat{A}\vec{u}^{(\ell)}\right)
\label{eqn:MultigridSmoother}
\end{equation}
and requires one sparse-matrix-vector product and one preconditioner solve in (\ref{eqn:SpMVPrec}). For the intergrid operations
\begin{xalignat}{2}
  \vec{u}^{(\ell)} &\mapsfrom \mat{R}_{\ell,\ell+1}\vec{u}^{(\ell+1)}
  \quad\text{(Restriction)},&
  \vec{u}^{(\ell)} &\mapsfrom \mat{P}_{\ell,\ell-1}\vec{u}^{(\ell-1)}
  \quad\text{(Prolongation)}
\end{xalignat}
we use a simple cell-average for the restriction and (piecewise) linear interpolation for prologation (both in the horizontal direction only), and we found that these methods are sufficient for scalable performance.
By carrying out the restriction at the beginning of the subroutine in Algorithm \ref{alg:VCycle} it is possible to fuse it with the first presmoothing step on the coarse levels. Apart from that fusing kernels has little potential for further gains in the multigrid algorithm. 

Recall that the condition number $\kappa_{\operatorname{fine}}$ of the fine grid problem is $\order(100-1000)$ independent of the horizontal resolution. The condition number of each subsequent coarse level is reduced by a factor $4$, i.e. the square of the relative grid spacings. We typically choose $L=5$ multigrid levels. Hence on the coarsest grid we have $\kappa_{\operatorname{coarse}}=4^{-(L-1)}\kappa_{\operatorname{fine}}$, i.e. the operator is well conditioned and the coarse grid equation can be solved by a small number (two turned out to be sufficient) of smoother iterations. This has already been confirmed by the detailed numerical experiments on CPU architectures in \cite{Mueller2013b}. For the particular model problem in (\ref{eqn:ModelEquation}) where $\omega(h)\propto h$ to accurately represent large scale atmospheric flow, both these iterative methods are algorithmically scalable, i.e. the number of iterations is independent of the mesh size $h$ and thus of the problem size. However, an additional benefit of multigrid solvers is their greater robustness with respect to variations in the model coefficients  \cite{Mueller2013b}.
\section{Implementation}\label{sec:Implementation}
In the following we describe the CUDA-C implementations of both the CG- and multigrid solvers discussed in the previous section. The code was written from scratch by the authors and we use the CUBLAS library for some level 1 BLAS operations as well as the GCL library \cite{bianco2013interface} for inter-GPU communication. The source code is made available under the LGPL 3 license and can be accessed as a git repository via the following link: \url{https://bitbucket.org/em459/ellipticsolvergpu}.
Further details on the single GPU implementation of our CG solver can be found in \cite{Mueller2013a}. 
\subsection{Memory throughput optimised implementation}
\label{sec:ImplementationMemory}
Due to the vertical dependency in the tridiagonal solver, which is used both
as the preconditioner in CG and as the smoother in multigrid, one thread
is assigned to each vertical column. To achieve optimal performance it is
crucial to coalesce access to global memory for all threads within one warp.
This is achieved by storing data contiguously in the horizontal ($x$-) direction.
As discussed in Sections 3.2 and 6.1 of \cite{Mueller2013a}, a three dimensional field $\vec{u}$ on one panel of a cubed sphere grid can be described as a collection of $n_x\times n_y$ vectors $\overline{\vec{u}}^{(i,j)}$, one for each horizontal cell $(i,j)\in[1,n_x]\times[1,n_y]$. Internally the field can be stored as a linear array of length $n_x\times n_y\times n_z$ defined by the mapping,
\begin{equation}
  u_{\linIdx(i,j,k)} = \overline{u}_k^{(i,j)}, \qquad \text{where}\quad
  \linIdx(i,j,k) \equiv n_x \cdot\left(n_z\cdot (j-1) + k\right)+(i-1).
  \label{eqn:MemoryMapSingleGPU}
\end{equation}
In the following we also assume that (at least on the finest multigrid levels) both $n_x$ and $n_y$ are multiples of 32. This further improves performance as global memory access is not only coalesced but also well-aligned.
We stress, however, that our approach can be generalised to more unstructured grids e.g. by the use of a space filling curve for numbering the horizontal grid cells.
\paragraph{Matrix-free implementation}
As in \cite{Mueller2013a} we use a matrix-free implementation, i.e. we recalculate the local matrix stencil whenever it is needed. In particular, the diagonal matrix $\mat{A}_{T,T'}$ and the tridiagonal matrix $\mat{A}_T$ in (\ref{eqn:TridiagonalPDE}) are given by
\begin{equation}
 \begin{aligned}
  \mat{A}_{T,T'} &= \alpha_{T,T'} \operatorname{diag}(\vec{d}), \\
  \mat{A}_{T} &= |T| \operatorname{diag}(\vec{a})-\alpha_T \operatorname{diag}(\vec{d}) + |T|\operatorname{tridiag}(-(\vec{b}+\vec{c}),\vec{b},\vec{c}).
 \end{aligned}
 \label{eqn:LocalMatrixStencil}
\end{equation}
The four vectors $\vec{a}$, $\vec{b}$, $\vec{c}$ and $\vec{d}$ have length $n_z$ and can be derived from the vertical stiffness- and mass- matrices in a tensor product representation of (\ref{eqn:ModelEquation}) \cite{Dedner2014}. As they do not differ from column to column they can be precomputed once for the entire grid. The coefficients $\alpha_{T,T'}$ and $\alpha_T$ are different for each horizontal grid cell $T$ (and depend on the multigrid level). However, they are scalars which can be computed for an entire vertical column with a small overhead as long as $n_z$ is sufficiently large (in atmospheric applications $n_z=\order(100)$, and we use $n_z=128$ throughout this work). This should be compared to a matrix-explicit implementation where seven matrix entries need to be loaded from memory per grid cell to carry out a sparse matrix-vector product. The matrix-free implementation significantly reduces global memory access, in particular if the vectors $\vec{a}$, $\vec{b}$, $\vec{c}$ and $\vec{d}$ are cached. Instead of $\order(7\times n_{\operatorname{horiz}}\times n_z)$ the storage requirements of the matrix are only $\order(4\times n_z)$ which also means that significantly larger problems can be solved on a single GPU.

At this point we mention that a very similar structure for the matrices $\mat{A}_{T,T'}$ and $\mat{A}_T$ arises when discretising more general equations of the form
\begin{equation*}
  -\omega^2 \vec{\nabla} \cdot(\mat{\alpha}\vec{\nabla} u) - \omega^2 \vec{\xi}\cdot \vec{\nabla} u + \beta u = f
\end{equation*}
as long as the ``profiles'' $\mat{\alpha}$, $\beta$ and $\vec{\xi}$ can be factorised into a parts which contain on the horizontal and vertical coordinates only. In this case the factors $|T|$, $\alpha_{T,T'}$ and $\alpha_{T}$ in (\ref{eqn:LocalMatrixStencil}) have to be replaced by more complicated expressions which can, however, still be evaluated only once per column. It is for this reason that we believe that the results in this article can be generalised easily to more complicated equations, this is discussed in more detail in \cite{Dedner2014}.

In addition, in the CG algorithm we reduce the amount of global memory access by fusing the two computationally most expensive kernels (SpMV and tridiagonal solve) with several of the level 1 BLAS operations. The following operations need to be carried out for the solvers and were implemented as CUDA-C kernels:
\paragraph{Conjugate Gradient}
\begin{enumerate}
  \item \textit{Sparse matrix-vector product} [Kernel \kSpMV]: Simultaneously calculate $\vec{u}\mapsfrom \vec{u}+\alpha\vec{p}$; $\vec{p}\mapsfrom \vec{z}+\beta\vec{p}$; $\vec{q}\mapsfrom \mat{A}\vec{z}+\beta\vec{q}$; $\sigma \mapsfrom \langle\vec{p},\vec{q}\rangle$ 
  \item \textit{Preconditioner (tridiagonal solve)} [Kernel \kTridiag]: Simultaneously calculate $\vec{r}\mapsfrom \vec{r}-\alpha\vec{q}$; $\vec{z}\mapsfrom\mat{M}^{-1}\vec{r}$; $||\vec{r}||\mapsfrom{\langle\vec{r},\vec{r}\rangle}$; $\kappa\mapsfrom\langle \vec{r},\vec{z}\rangle $
\end{enumerate}
\paragraph{Multigrid}
\begin{enumerate}
  \item \textit{Smoother} [Kernel \kSmooth]: $\vec{u}^{(\ell)} \mapsfrom \vec{u}^{(\ell)} + \rhorelax \mat{M}^{-1}(\vec{f}^{(\ell)}-\mat{A}\vec{u}^{(\ell)})$; to avoid a race condition in $\vec{u}^{(\ell)}$, this operation is split up into two kernels, with the residual calculation and forward sweep of the tridiagonal solver in the first and the backward sweep and axpy-like update of $u$ in the second.
  \item \textit{Residual calculation} [Kernel \kRes]: $\vec{r}^{(\ell)}\mapsfrom \vec{f}^{(\ell)}-\mat{A}\vec{u}^{(\ell)}$
  \item \textit{Interleaved (fused) restriction and smoother} [Kernel \kResSmooth]: Simultaneously calculate $\vec{f}^{(\ell)}\mapsfrom\mat{R}_{\ell,\ell+1} \vec{r}^{(\ell+1)}$, $\vec{u}^{(\ell)} \mapsfrom \rhorelax \mat{M}^{-1}\vec{f}^{(\ell)}$
  \item \textit{Prologation} [Kernel \kProlongate]:  $\vec{u}^{(\ell)}=\vec{u}^{(\ell)}+\mat{P}_{\ell,\ell-1}\vec{u}^{(\ell-1)}$
\end{enumerate}
Each of the kernels requires a single iteration over the grid. For the multigrid solver the first presmoothing step on the coarse grids (on which the initial solution is initialised to zero) has been fused with the residual calculation to reduce the number of accesses to the vector $\vec{f}^{(\ell)}$.

In addition, a small number of level 1 BLAS operations still needs to be carried out for global reductions. For example, the global reductions in the interleaved CG kernels are implemented by each thread summing up the values in a vertical column into a two-dimensional field, which is then summed by a cuBLAS dot product with a field that is set to 1 in the whole two dimensional domain. For simplicity the calculation of the norm of the residual on the finest multigrid level was also implemented via a 3D cuBLAS norm instead of fusing it with the corresponding kernel (note that this norm calculation is also not necessary if a fixed number of V-cycles is carried out). While there is further potential for (small) additional speedups, we found that the cost of these level 1 BLAS operations is negligible (to see this, compare the last two rows in Tables \ref{tab:TimingBreakdownCG} and \ref{tab:TimingBreakdownMultigrid} below) and not worth the effort.
\subsection{Multi-GPU implementation}
To parallelise the solvers across several GPUs we split the horizontal domain
into equal square parts, such that each GPU is responsible for a quadratic
subdomain. Consistency between neighbouring domains is guaranteed by
exchanging halo data when necessary. For this we used the Generic
Communication Library (GCL) \cite{bianco2013interface}. The extension of this approach to more general partitionings and different (semi-) structured horizontal grids is possible. This is discussed in \cite{Dedner2014} where we discuss a parallel CPU implementation of the tensor-product multigrid solver on an icosahedral grid.

\begin{figure}
 \begin{center}
   \includegraphics[width=0.4\linewidth]{\figdir/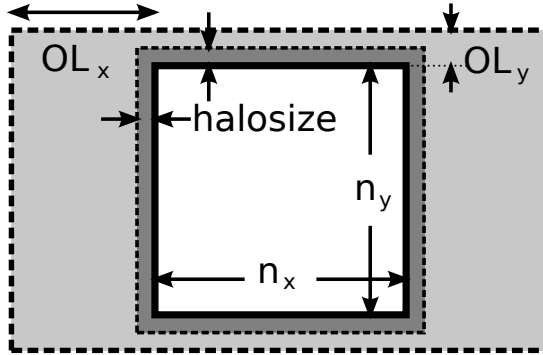}
  \caption{Data layout of the local subdomain on one processor. Interior degrees of freedom $(i,j)\in [1,n_x]\times [1,n_y]$ ``owned'' by the processor are shown in white, the halo is shown in dark gray and extra padding space in light gray. We set $\OL_x=32$ and $\OL_y=\halosize$ to guarantee aligned memory access in the $x$-direction.}
  \label{fig:Domain}
 \end{center}
\end{figure}

In addition to the interior degrees of freedom with horizontal indices $(i,j)\in[1,n_x]\times[1,n_y]$ a halo of cells of width $\halosize=1$ is stored on each GPU. 
To avoid unaligned memory access, which we found can reduce performance by as much as $30\%$, we pad data by a total amount of $\OL_x=32$ in the $x-$ direction, and set $\OL_y=\halosize$ in the $y-$ direction (in this direction padding is not necessary). Then the linear mapping (\ref{eqn:MemoryMapSingleGPU}) is modified to 
\begin{equation}
  \hat{\linIdx}(i,j,k) \equiv (n_x+2\OL_x)\cdot\left(n_z\cdot(j-1-\OL_y)\right)+ (i-1-\OL_x)
\label{eqn:xContiguousLayout}
\end{equation}
with $(i,j) \in [1-\OL_x,n_x+\OL_x-1]\times [1-\OL_y,n_y+\OL_y-1]$; the local domain is shown in Figure \ref{fig:Domain}.
We stress, however, that only degrees of freedom on the halo cells are exchanged between processors i.e. the padding does not increase the amount of data that is sent over the network. In the GCL this can be achieved by registering the appropriate degrees of freedom in the halo exchanger object
\begin{verbatim}
    he->add_halo<0>(halosize, halosize, OL_X, NX+OL_X-1, NX+2*OL_X);
    he->add_halo<1>(halosize, halosize, OL_Y, NY+OL_Y-1, NY+2*OL_Y);
    he->add_halo<2>(0, 0, 0, NZ-1, NZ);
\end{verbatim}
where \texttt{he} is an instantiation of the \textit{uniform type} halo structure interface class \verb!GCL::halo_exchange_dynamic_ut!.
\section{Theoretical performance analysis}\label{sec:PerformanceAnalysis}
\subsection{Floating point operations and memory transfer costs}
\label{sec:PerformanceAnalysisSingleGPU}
The number of floating point operations (FLOPs) and memory references per grid cell is shown for the CG and multigrid algorithms in Table \ref{tab:KernelTable}.
 The total number of operations on all 5 multigrid levels (penultimate row in Table \ref{tab:KernelTable}, right) was obtained by adding up the number of operations on all levels and dividing by the size of the fine grid. We assume that 1 pre- and post-smoothing step is applied on each level and the coarse grid problem is solved by two smoother iterations. The residual norm is only calculated on the fine level to test for convergence. 

The column labelled with ``\Mem'' shows the number of global memory references without any caching. On the other hand the ``\MemC'' column shows the corresponding number assuming optimal caching, i.e. we assume that data is only loaded from global memory once per kernel call. As our implementation is matrix-free (and we assume that the vectors $\vec{a}$, $\vec{b}$, $\vec{c}$ and $\vec{d}$ in (\ref{eqn:LocalMatrixStencil}) are always cached), there are no costs associated with reading the local stencil from global memory. Consider, for example, the sparse matrix vector product $\vec{y}\mapsfrom \mat{A}\vec{x}$. Without caching, the value of the vector $\vec{x}$ in each cell and its six direct neighbours needs to be read from memory and one value is written back to $\vec{y}$, resulting in a total of 8 memory references. With perfect caching each entry in $\vec{x}$ only has to be read from global memory once and the total number of memory references is reduced to 2 per grid cell. How much caching can actually be achieved is hard to predict, so these two values should be interpreted as upper and lower bounds. As explained in Section \ref{sec:SingleGPUPerformance} we always use the lower value (i.e. the number in the ``\MemC'' column) for our estimates of the achieved global memory bandwidth. The quantity which is constructed in this way is commonly know as the ``useful bandwidth'' since it does not include any spurious memory traffic which is not required by the algorithm (for example data which is read twice due to poor caching). 

In the multigrid solver the grid size is reduced by a factor 4 with each coarsening step and as a result most of the computational work is concentrated on the finest grid level. To demonstrate this we also list the number of operations per grid cell on the fine level only in the last row of Table \ref{tab:KernelTable}, right. In practise, once the size of the horizontal grid drops below a certain threshold, the GPU might not be utilised efficiently and the actual runtime is reduced by less than a factor four on subsequent levels. Our measurements (see Table \ref{tab:SingleGPUTiming}) show that calculations on the fine grid account for most of the total computational cost and that the cost reduction factor is close to four as long as the local domain is larger that $64\times 64$ (see Figure \ref{fig:MultigridBreakdownLogarithmic} (left)). The impact of inter-GPU communications, which is also more significant on the coarser grids, will be discussed below.

For the CG solver, the number of FLOPs is only $3.6\times$ larger than the minimal number of memory references, for the multigrid solver the corresponding factor is $5\times$, and we conclude that both algorithms are clearly memory bound on a GPU. By comparing the number of memory references in the two algorithms, the theoretically expected time per iteration is $2\times$ to $3\times$ larger for the multigrid solver where the exact ratio depends on the cache efficiency.

It is worth counting the additional number of global memory accesses which would be required in a matrix-explicit code: to read the matrix from memory requires 10 memory accesses for the CG solver and 42.9 for the multigrid algorithm (the full seven point stencil is required to evaluate the sparse matrix-vector product, but only the couplings to the cell above and below are required in the tridiagonal solver). This should be compared to the minimal number of memory references shown for the matrix-free code in Table \ref{tab:KernelTable} which is 15 for CG and 29.6 for multigrid.
\begin{table}
\renewcommand{\arraystretch}{1.25}
\begin{center}
\begin{minipage}{0.48\linewidth}
\begin{center}
(Fused) CG\\
\begin{tabular}{|l|rrr|}
\hline
& \FLOPs & \Mem & \MemC \\\hline\hline
\kSpMV &32 &12 &6\\
\kTridiag &22 &12 &9\\
\hline
\textbf{Total} &\textbf{ 54} &\textbf{ 24} &\textbf{ 15}\\
\hline
\end{tabular}
\end{center}
\end{minipage}
\hfill
\begin{minipage}{0.48\linewidth}
\begin{center}
Multigrid\\
\begin{tabular}{|l|rrr|}
\hline
& \FLOPs & \Mem & \MemC \\\hline\hline
\kSmooth &37 &17 &8\\
\kResSmooth &17 &12 &6\\
\kRes &23 &9 &3\\
\kProlongate &6 &5 &3\\
\kResNorm &2 &1 &1\\
\hline
\textbf{Total} &\textbf{149.4} &\textbf{67.2} &\textbf{29.6}\\
\hline(fine level only) &(122) &( 53) &( 23)\\
\hline
\end{tabular}
\end{center}
\end{minipage}
\caption{Number of FLOPs and memory references per grid cell for the kernels in the (fused) CG algorithm (left) and the multigrid solver (right); see Section \ref{sec:ImplementationMemory} and Algorithm \ref{alg:VCycle} for a definition of the individual kernels.}
\label{tab:KernelTable}
\end{center}
\end{table}

\subsection{Parallel communications between GPUs}
\label{sec:PerformanceAnalysisComm}
In the CG solver a halo exchange is required after each call to the (fused) tridiagonal solver kernel. In the multigrid algorithm halos need to be exchanged after each kernel launch (with the exception of the residual calculation). Denoting the number of halo exchanges per iteration by $n_{\operatorname{halo}}$ and the minimal number of memory references by $n_{\MemC}$, on a given level with a local problem size of $n_x\times n_y\times n_z$ (where we always implicitly assume that $n_x=n_y$), the ratio between the communication time and calculation time is given by
\begin{equation}
  \rho = \frac{2(n_x+n_y)n_z\times \operatorname{sizeof}(\texttt{double})\times \halosize\times n_{\operatorname{halo}}}{n_xn_yn_z \times n_{\MemC}}\times \frac{\BWmem}{\BWMPI} \equiv R \times\frac{\BWmem}{\BWMPI}.
\label{eqn:CommCalcRatio}
\end{equation}
This ratio decreases as $\propto (n_x+n_y)/(n_x\cdot n_y) \propto 1/n_x$ as the local domain size $n_x$ increases. We assume that the global memory bandwidth $\BWmem$ is about two orders of magnitude larger than the network bandwidth $\BWMPI$ required for communication between different GPUs (the exact ratio between the bandwidths will be quantified in more detail in Section \ref{sec:MultiGridParallelOverhead}). For the multigrid solver the amount of exchanged data and the number of memory references have to be summed over all levels to calculate the ratio $R$ in (\ref{eqn:CommCalcRatio}). 

In Table \ref{tab:CommCalcRatio} the ratio $R$ is shown for all solvers. For the multigrid algorithm we also distinguish between the full level hierarchy and the finest level only. While (based on our estimate $\BWmem/\BWMPI\approx 100$) the resulting ratio $\rho$ is at the order of $10\%$ or less for the CG solver and on the finest multigrid level (for local problem sizes larger than $256\times256\times128$), it is approximately four times larger for the multigrid solver. The main reason for this is that three halo exchanges are required on each level instead of one for the CG solver. In addition on the coarser levels the ratio between communications and calculations is a factor $n_x/n^{(\ell)}_x>1$ larger than on the finest level, where $n^{(\ell)}_x=n^{(\ell)}_y$ is the problem size on level $\ell$. With $L=5$ multigrid levels this corresponds to an increase by a factor $2^{L-1}=16$ on the coarsest level.
For both reasons we expect the overhead from MPI communications to be larger for the multigrid solver and this is confirmed by our numerical experiments in Section \ref{sec:MultiGridParallelOverhead}.
\begin{table}
\begin{center}
\renewcommand{\arraystretch}{1.25}
\begin{tabular}{|l|rrrr|}
\hline
 & \multicolumn{4}{|c|}{Problem size $n_x$}\\
Solver & $128$ & $256$ & $512$ & $768$ \\\hline\hline
(Fused) CG
&20.8
&10.4
&5.2
&3.5
\\
Multigrid
&80.5
&40.2
&20.1
&13.4
\\
Multigrid (fine level)
&54.3
&27.2
&13.6
&9.1
\\
\hline
\end{tabular}
\caption{Theoretical ratio between MPI communication- and calculation- times as defined in (\ref{eqn:CommCalcRatio}). The table shows $R$ multiplied by $10^4$ for different local horizontal problem sizes $n_x=n_y$ with $\halosize=1$ and double precision arithmetic ($\text{sizeof(\texttt{double})}=8$).}
\label{tab:CommCalcRatio}
\end{center}
\end{table}
\section{Results}\label{sec:Results}
In the following we report on detailed numerical experiments with our solvers and quantify both their algorithmic- and computational performance on up to 16384 nVidia GPUs.
\paragraph{Hardware and compilers}
Most GPU results shown in this section were obtained on the Cray XK7 Titan supercomputer at Oak Ridge National Lab. 
According to the June 2014 release of the \texttt{top500.org} list \cite{top500.org} this machine is currently ranked as the second fastest computer in the world and consists of 18,688 nodes in total. Each node contains one 16-core 2.2GHz AMD Opteron(TM) 6274 (Interlagos) CPU and one nVidia Tesla K20X card with a GK110 Kepler GPU (compute capability 3.5). The nodes are linked with a Cray Gemini interconnect in a torus topology. Each GPU has 2688 cores, which are organised into 15 streaming multiprocessors of 192 cores each. The K20X card has 6 GB of global memory, in addition to 1536KB L2 cache and 48KB+16KB configurable L1 cache/shared memory. The theoretical peak FLOP rate of a single GPU is 1.31 \TFLOPs\ and the theoretical peak global memory bandwidth is $250\GBytes$ \cite{TeslaDatasheet}, resulting in a theoretical peak floating point performance of 27 \PFLOPs\ (combined CPU and GPU performance); the LinPACK benchmark performance is quoted as 17.59 \PFLOPs\ \cite{top500.org}. The code was compiled with release 5.0 (V0.2.1221) of the nVidia nvcc compiler and version 4.7.2 of the gnu c++ compiler; a vendor optimised MPICH2 implementation that supports GPUDirect was used for internode communications.

To study the dependency of the performance on the GPU hardware we also carried out a number of runs on up to 64 GPUs of the EMERALD GPU cluster hosted at Rutherford Appleton Lab in the UK. This cluster consists of 372 NVidia M2090 cards with Fermi GPUs (compute capability 3.0), which are organised into 60 nodes with 3 GPUs each and 24 nodes with 8 GPUs each. With 0.67\TFLOPs\ the double precision floating point performance is half as large as for the Kepler GK110 cards in Titan. The global memory bandwidth is $177\GBytes$. Each node contains two 6-core Intel X5650 Xeon CPUs and a QDR Infiniband network is used to link the nodes; the MVAPICH2 implementation was used for MPI communications.

To quantify the performance of our algorithms on traditional CPU hardware we compared the GPU code to a bespoke Fortran implementation of the multigrid solver which is based on the MPI distributed memory programming model. The Fortran code was run on up to 65536 cores of HECToR, the UK's national supercomputer which is hosted and managed by the Edinburgh Parallel Computing Centre
(EPCC). The performance and scalability of this Fortran code is discussed in more detail in \cite{Mueller2013b}. HECToR is a Cray XE6 supercomputer consisting of 2816 compute nodes connected by a Cray Gemini interconnect. Each node contains two AMD Opteron (Interlagos, model 6276, 2.3 GHz) processors with 16 cores \cite{OpteronDatasheet}, and we adjusted the problem size such that it is the same on one GPU and on one CPU, i.e. we always carried out a socket-to-socket comparison. 
\paragraph{Problem- and solver- parameters}
For a given problem size, which in the following is defined by the number of horizontal grid cells $n_x$ in one direction, the parameters $\omega(h)$ and $\lambda(h)$ in (\ref{eqn:ModelEquation}) were adjusted to physically realistic values, for the exact values compare Table \ref{tab:Problemsizes} to Table 1 in \cite{Mueller2013b}. In particular $\omega(h)$ decreases linearly with increasing model resolution as described in (\ref{eqn:OmegaNumerical}), which implies that the condition number of the discretised equations does not increase with problem size, and hence the number of iterations is roughly independent of $n_x$. As estimated above, the condition number is $\order(100-1000)$. The number of vertical grid levels was kept fixed at $n_z=128$.

We use five multigrid levels in all cases and solve the coarse grid problem by applying two smoother iterations and use one pre- and post-smoothing step on all multigrid levels. The overrelaxation parameter in the block-Jacobi smoother was set to $\rhorelax=2/3$, which is the optimal value for the two dimensional Poisson equation.
\subsection{Single GPU performance}\label{sec:SingleGPUPerformance}
The performance of CG and of the multigrid solver is shown in Table \ref{tab:SingleGPUTiming} for different problem sizes $n_x$. The largest problem that could be solved with the CG algorithm has $768\times768\times128=7.55\cdot10^{7}$ degrees of freedom, whereas for the multigrid solver the largest problem that fits into memory has $512\times 512\times 128=3.36\cdot 10^{7}$ unknowns. 
On the GPU the horizontal thread layout was chosen such that each block consists of 64 threads in the (memory contiguous) $x$- direction and 2 threads in the $y$- direction. We iterate in both cases until the residual has been reduced by five orders of magnitude, i.e. we use $\varepsilon=10^{-5}$ in (\ref{eqn:epsilonTolerance}). The number of iterations is almost 8 times smaller for the multigrid solves. The numbers depend only weakly on the problem size, confirming the good algorithmic scalability of both solvers. For the $n_x=512$ problem, one iteration of the multigrid solver is three times as expensive as a CG iteration which is in line with the prediction based on the number of memory references in Section \ref{sec:PerformanceAnalysisSingleGPU}. This ratio deteriorates for smaller problem sizes because as the problem size decreases, the GPU will be underutilised on the coarse grid levels, see also Figure \ref{fig:MultigridBreakdownLogarithmic}.

At this point it is worth recalling the significant reduction in runtime which is achieved by not storing the matrix explicitly. In \cite{Mueller2013a} the performance of the matrix-free fused CG solver is compared to an implementation based on the CUSparse library which stores the matrix in compressed sparse row (CSR) storage format. For a $256\times256\times128$ problem on a single Fermi M2090 GPU a speedup of a factor $4.6\times$ is achieved by recalculating the local matrix stencil on-the-fly instead of using the CSR implementation.

In Table \ref{tab:SingleGPUTiming} we also quantify the floating point performance and the percentage of the peak global memory bandwidth which can be utilised by our solvers. For this we use the figues in Table \ref{tab:KernelTable} and divide them by the measured time per iteration in the first row of Table \ref{tab:SingleGPUTiming}. In all cases our calculation is based on the minimal number of memory accesses, i.e. the number in the ``\MemC'' column, i.e. we calculate the ``useful bandwidth''. For a memory bound application a bandwidth of $100\%$ would correspond to perfect caching. It is very hard to achieve this theoretical upper bound in practice.

We stress that the total solution time in the last row includes the time for copying the right hand side from the host to the device and for copying the final solution back to the host. This time is listed separately in Table \ref{tab:SingleGPUTiming} together with the time for transposing the fields from a $z$-contiguous data format on the host to the $x$-contiguous format in (\ref{eqn:xContiguousLayout}) on the GPU. For optimal efficiency the transposition was implemented on the GPU by adapting the algorithm described by Mark Harris \cite{Harris2013}. Looking at the $512$ problem again, the multigrid solver converges twice as fast as the CG iteration. This speedup is reduced for smaller problem sizes but still $1.6\times$ for the smallest problem we considered. In summary this stresses the point made at the beginning of this article: the highest performance gains can be achieved by choosing the algorithmically most efficient solver even if it is more expensive and computationally less efficient in a single iteration and has a slightly worse parallel efficiency.
\begin{table}
 \begin{center}
\renewcommand{\arraystretch}{1.25}
\begin{tabular}{|l|rrrr|rrr|}\hline
&\multicolumn{4}{|l|}{(Fused) CG}&\multicolumn{3}{|l|}{Multigrid}\\
&\multicolumn{4}{|c|}{Problem size $n_x$}&\multicolumn{3}{|c|}{Problem size $n_x$}\\
 & 128 & 256 & 512 & 768 & 128 & 256 & 512\\\hline\hline
$t_{\operatorname{iter}}$ & $ 2.8$ & $ 7.5$ & $29.7$ & $65.0$ & $12.8$ & $27.3$ & $88.8$\\
GFLOPs & $41.0$ & $60.1$ & $60.9$ & $62.7$ & $24.6$ & $45.9$ & $56.4$\\
percentage of peak BW & $36.4\%$ & $53.4\%$ & $54.2\%$ & $55.7\%$ & $15.6\%$ & $29.1\%$ & $35.8\%$\\
\# iterations & $70$ & $62$ & $59$ & $58$ & $9$ & $8$ & $8$\\
\hline
\textbf{$t_{\operatorname{MemCpy+transpose}}$} & $19.4$ & $63.5$ & $228.4$ & $494.0$ & $19.6$ & $64.3$ & $229.5$\\
\hline
\textbf{total solution time} & $\boldsymbol{217.1}$ & $\boldsymbol{543.7}$ & $\boldsymbol{2029.0}$ & $\boldsymbol{4365.0}$ & $\boldsymbol{135.5}$ & $\boldsymbol{285.7}$ & $\boldsymbol{949.9}$\\
\hline
\end{tabular}  
\caption{
Time per iteration $t_{\operatorname{iter}}$, number of iterations and total solution time for different problem sizes $n_x$ and solvers. 
The total solution time includes the host-device memory transfer and data transposition time, which is listed separately as $t_{\operatorname{MemCpy+transpose}}$. All times are given in milliseconds. The floating point performance and percentage of theoretical peak global memory bandwidth are calculated based on $t_{\operatorname{iter}}$ and the numbers in Table \ref{tab:KernelTable}, see main text for details.}
  \label{tab:SingleGPUTiming}
 \end{center}
\end{table}
\subsection{Robustness}\label{sec:Robustness}
\begin{figure}
  \begin{center}
    \includegraphics[width=0.45\linewidth]{\figdir/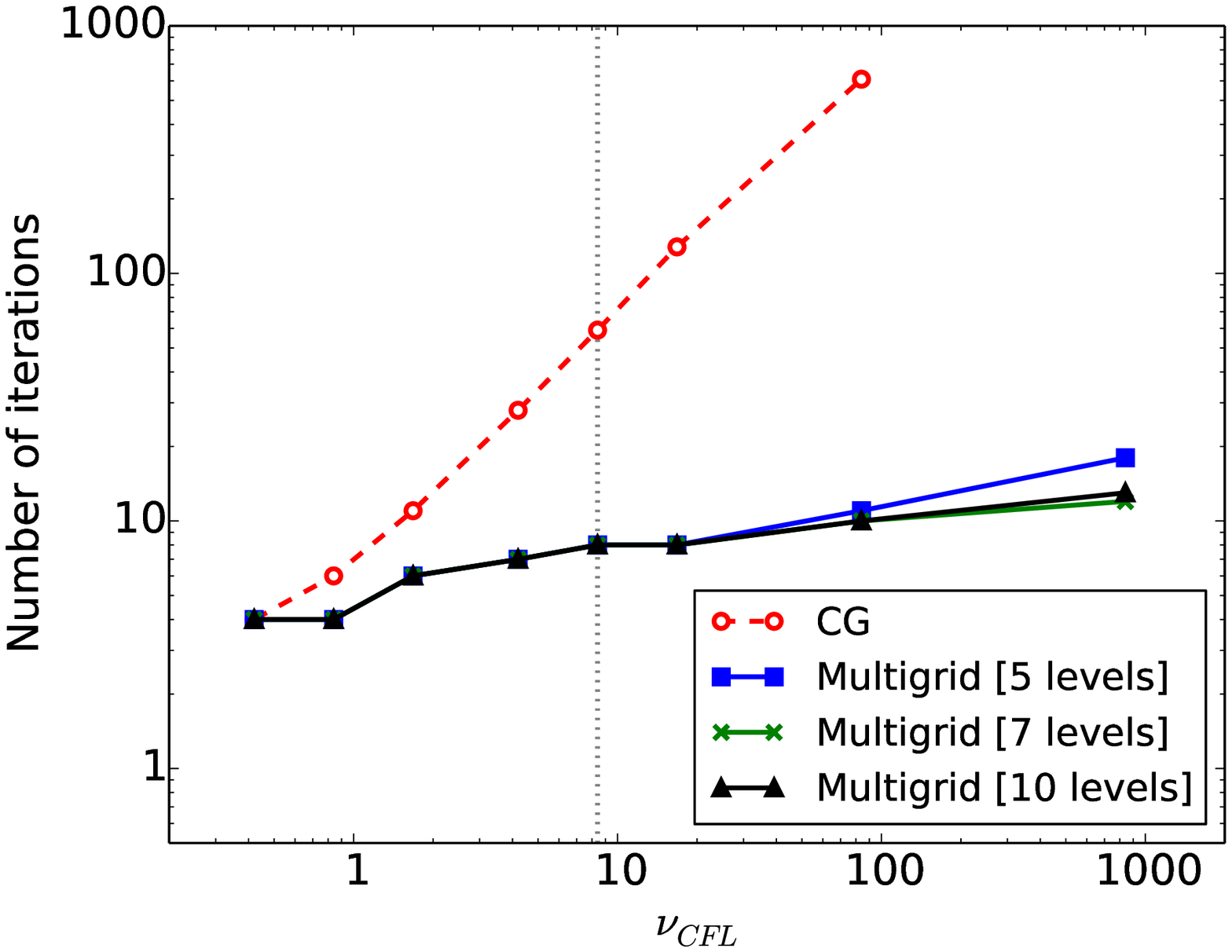}
    \includegraphics[width=0.45\linewidth]{\figdir/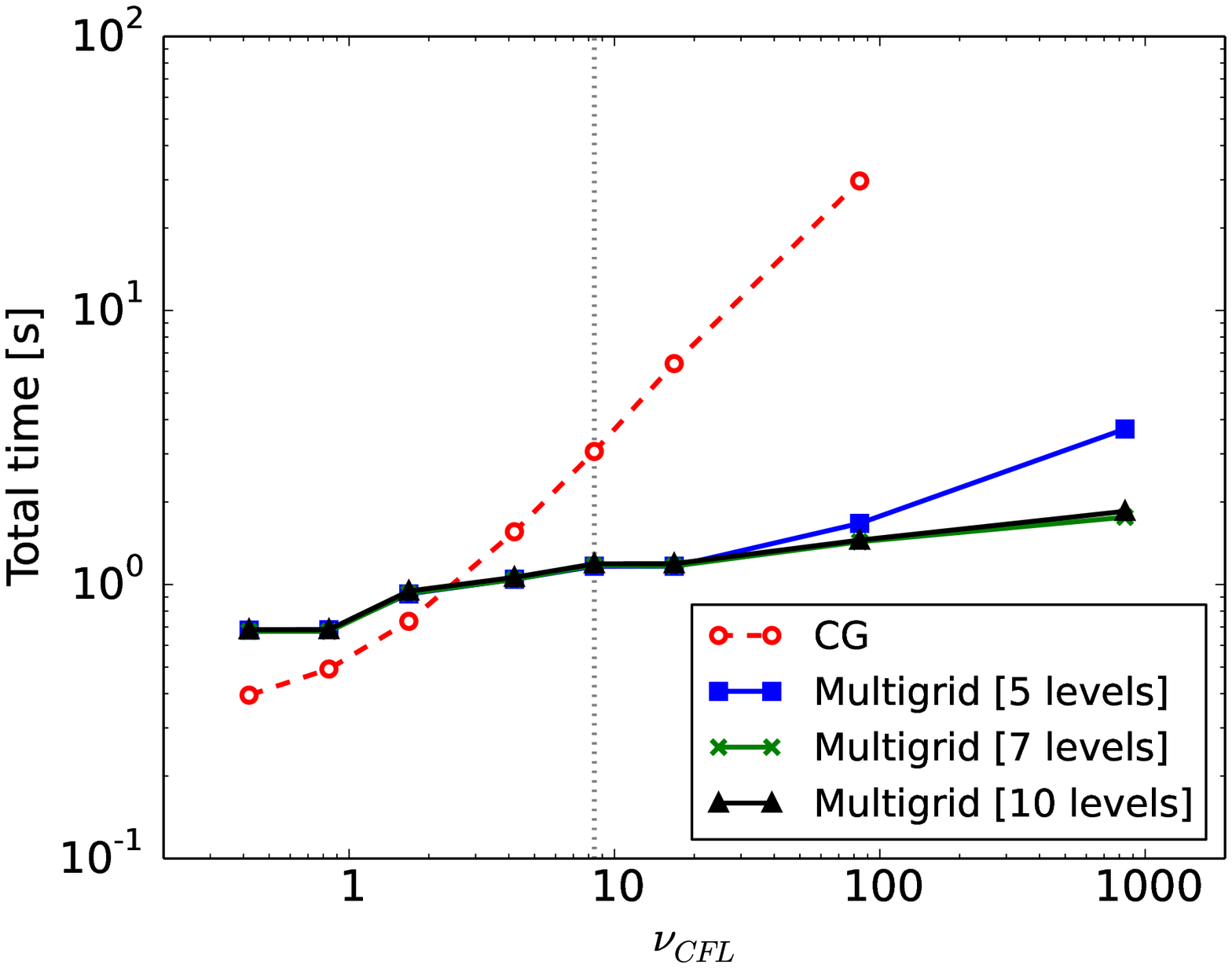}\\
    \caption{Number of iterations (left) and total solution time (right) for different solvers and a range of CFL numbers. The problem size was fixed to $512\times 512 \times 128$ and all runs were carried out on one M2090 GPU of the EMERALD cluster. The dashed vertical line marks the CFL number used for numerical experiments in the rest of this article.}
  \label{fig:Robustness}
  \end{center}
\end{figure}

For Krylov solvers the algorithmic performance, i.e. the number of iterations to reduce the relative residual below a certain threshold, depends on the condition number $\kappa(h)$ of the preconditioned elliptic operator. As discussed above, in the case of vertical line relaxation, $\kappa(h)$ only depends on the ratio of $\omega(h)$ and the grid spacing $h$; the size of $\kappa(h)$ is estimated in (\ref{eqn:ConditionNumber}). The dimensionless quantity $\omega(h)$ is proportional to the time step size (see (\ref{eqn:omegahdependence})) and hence $2\omega(h)/h=\nu_{CFL}$ is the horizontal CFL number. As already shown in \cite{Mueller2013b}, the tensor-product multigrid method converges in a constant number of iterations for our model problem, independent of the condition number. While for most numerical experiments presented in this paper we fixed $\nu_{CFL}=8.4$, here we also study the algorithmic performance for other choices of $\nu_{CFL}$ for fixed horizontal resolution. In the limit $\Delta t\rightarrow \infty$, $\nu_{CFL}\rightarrow\infty$ one would recover the Laplace equation from (\ref{eqn:ModelEquation}). In Fig. \ref{fig:Robustness} both the number of iterations and the total solution time are shown for different values of $\nu_{CFL}$. In addition to the results for 5 multigrid levels, which were used for the numerical experiments in the rest of this paper, we also show results for 7 and 10 multigrid levels. 
As expected, since $\kappa(h)\propto \nu_{CFL}^2$ for $\nu_{CFL}\gg 1$, the number of CG iterations increases linearly with $\nu_{CFL}$ for the CG solver. In terms of the total solution time CG only becomes competitive for small CFL numbers.

The multigrid algorithm is significantly more robust. The number of iterations and the total solution time depends only weakly on the CFL number.
As the CFL number increases, the coarse grid problem becomes less well conditioned. Depending on the number of levels, it may be necessary to increase the number of smoother iterations on the coarsest grid (or use a different coarse grid solver, such as preconditioned CG). We found that for the 10-level method 2 smoother iterations are sufficient to solve the coarse grid problem up to a value of $\nu_{CFL}=840$ without any significant increase in the number of iterations or in the total solution time. For the 7-level method  it was necessary to increase the number of smoother iterations on the coarsest level slightly for $\nu_{CFL} > 16.8$. To maintain robustness, 5 and 15 smoother steps were sufficient for $\nu_{CFL} = 84$ and $\nu_{CFL} = 840$, respectively. For the 5-level method, still 2 smoother steps on the coarsest level suffice up to $\nu_{CFL} =16.8$, but the number has to increase significantly faster for larger CFL numbers, namely to $30$, for $\nu_{CFL} = 84$, and to $150$, for $\nu_{CFL}$ = 840, leading to a slightly faster increase of the number of iterations and of the total solution time for the 5-level method. Nevertheless, with these modifications the multigrid solver is robust over a very wide range of CFL numbers.
\subsection{Communication overhead between GPUs and multigrid performance}\label{sec:MultiGridParallelOverhead}
To identify the bottlenecks of the multigrid algorithm and also to study the impact of parallel communications, the time per iteration was broken down into the time spent in the individual kernels both for a 1 GPU run and for a 64 GPU run with identical local grid size. For the (fused) SpMV and preconditioner kernels these times are shown in Table \ref{tab:TimingBreakdownCG} for a local $512\times 512\times 128$ grid. The total time per iteration increases by around $10\%$ due to the halo exchange, this should be compared to the theoretically predicted increase of around $5\%$ according to Table \ref{tab:CommCalcRatio}. The corresponding results are shown for the multigrid solver in Table \ref{tab:TimingBreakdownMultigrid}. Here the time per iteration grows by nearly $40\%$ when going from one to 64 GPUs, while the theoretical analysis predicts a $20\%$ increase.

For the theoretical estimates of the communication/calculation ratio $R$ in (\ref{eqn:CommCalcRatio}) we quantify the memory and inter-GPU bandwidth as follows: the peak global memory bandwidth of the K20X card is $\BWmem=250GB/s$, and as demonstrated in Section \ref{sec:SingleGPUPerformance}, our solvers can typically utilise at the order of $30\%-50\%$ of this peak value on a single GPU (see Table \ref{tab:SingleGPUTiming}). We measured the communication bandwidth by carrying out 1000 halo exchanges on 64 GPUs and obtained $\BWMPI\approx1GB/s$, which implies that $\BWmem/\BWMPI=\order(100)$.
\begin{table}
 \begin{center}
\renewcommand{\arraystretch}{1.25}
\begin{tabular}{|l|r|r|}
\hline & 1 GPU & 64 GPUs\\
\hline\hline
(Fused) SpMV  & $14.1$  & $14.2$ \\
(Fused) Tridiag  & $15.6$  & $18.0$ \\
\hline
\textbf{Total [kernels]}
 & \textbf{29.7}  & \textbf{32.2} \\
\hline
Total [iteration]
 & $29.8$  & $32.5$ \\
\hline
\end{tabular}
  \caption{Breakdown of the (fused) CG solver time per iteration on a $512\times 512\times 128$ grid for 1 and 64 GPUs [all times in milliseconds].}
  \label{tab:TimingBreakdownCG}
 \end{center}
\end{table}

\begin{figure}
 \begin{center}
  \includegraphics[width=0.45\linewidth]{\figdir/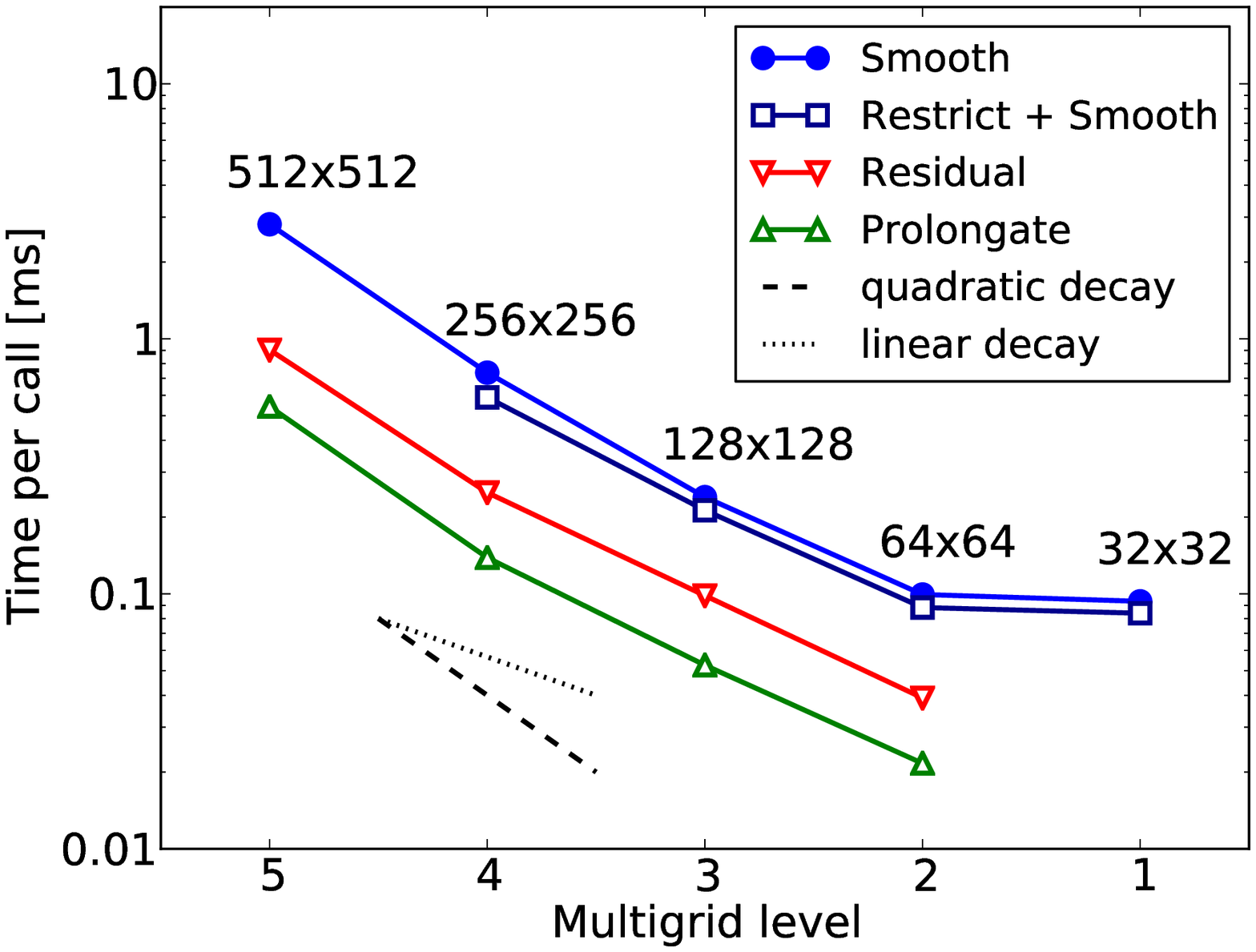}
  \includegraphics[width=0.45\linewidth]{\figdir/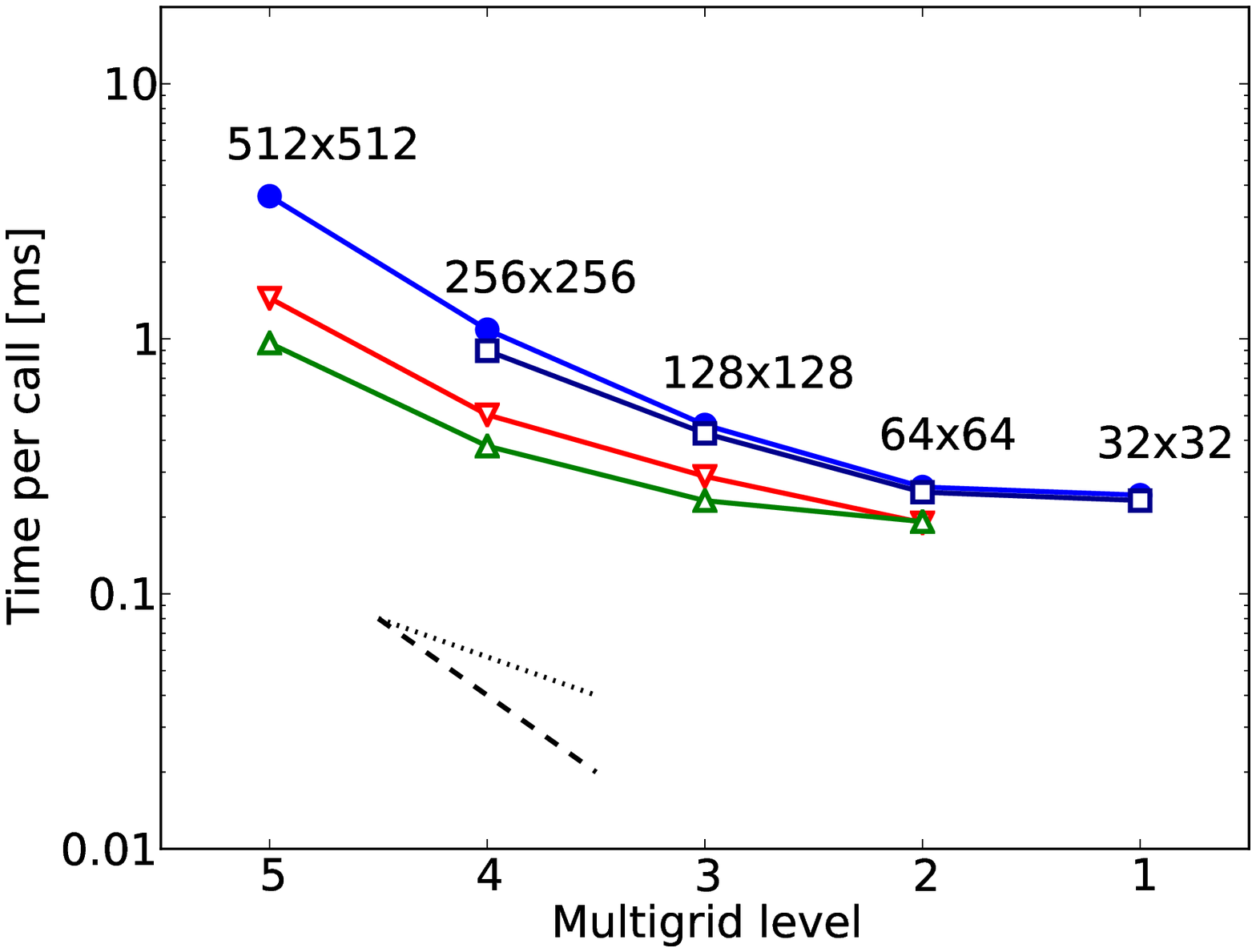}
  \caption{Breakdown of the multigrid solver time on a $512\times 512\times 128$ grid on
  1 (left) and 64 GPUs (right). The horizontal grid size is shown separately for each multigrid level.}
  \label{fig:MultigridBreakdownLogarithmic}
 \end{center}
\end{figure}

Looking at the time spent on different multigrid levels, which is also plotted on a logarithmic scale in Figure \ref{fig:MultigridBreakdownLogarithmic}, we see that part of this poor scaling can be attributed to a worse calculation/communication ratio on the coarser levels. On a single GPU the times decrease by roughly a factor of 4 from level to level as expected, until a horizontal problem size of $64\times 64$ is reached. Beyond this point the costs do not decrease further as the GPU is underutilised. However, on 64 GPUs the cost per level is reduced by less than a factor 4 on all levels due to the worse communication and calculation ratio.
\begin{table}
 \begin{center}
\renewcommand{\arraystretch}{1.25}
\begin{tabular}{|l|rrrrr|rrrrr|}
\hline & \multicolumn{5}{|l|}{1 GPU}  & \multicolumn{5}{|l|}{64 GPUs} \\
 & \multicolumn{5}{|c|}{Multigrid level} & \multicolumn{5}{|c|}{Multigrid level}\\
kernel & 5 & 4 & 3 & 2 & 1 & 5 & 4 & 3 & 2 & 1\\
\hline\hline
Smooth  & $45.0$ & $ 5.9$ & $ 1.9$ & $ 0.8$ & $ 0.7$ & $50.7$ & $ 7.6$ & $ 3.2$ & $ 1.8$ & $ 1.7$\\
ResSmooth  & --- & $ 4.7$ & $ 1.7$ & $ 0.7$ & $ 0.7$ & --- & $ 6.3$ & $ 3.0$ & $ 1.8$ & $ 1.6$\\
Residual  & $15.4$ & $ 2.0$ & $ 0.8$ & $ 0.3$ & --- & $21.7$ & $ 3.5$ & $ 2.0$ & $ 1.3$ & ---\\
Prolongate  & $ 4.3$ & $ 1.1$ & $ 0.4$ & $ 0.2$ & --- & $ 6.7$ & $ 2.7$ & $ 1.6$ & $ 1.3$ & ---\\
\hline
\textbf{Total [kernels]}
 & \textbf{64.7} & \textbf{13.7} & \textbf{ 4.8} & \textbf{ 2.0} & \textbf{ 1.4} & \textbf{79.1} & \textbf{20.1} & \textbf{ 9.8} & \textbf{ 6.3} & \textbf{ 3.3}\\
\hline
Total [iteration]
 & \multicolumn{5}{|c|}{$88.8$}  & \multicolumn{5}{|c|}{$122.0$} \\
\hline
\end{tabular}
  \caption{Breakdown of the multigrid solver time per iteration on a $512\times 512\times 128$ grid for 1 and 64 GPUs [all times in milliseconds].}
  \label{tab:TimingBreakdownMultigrid}
 \end{center}
\end{table}

\subsection{Massively parallel scaling on GPU and CPU clusters}\label{sec:WeakScaling}
\begin{figure}
 \begin{center}
  \includegraphics[width=0.45\linewidth]{\figdir/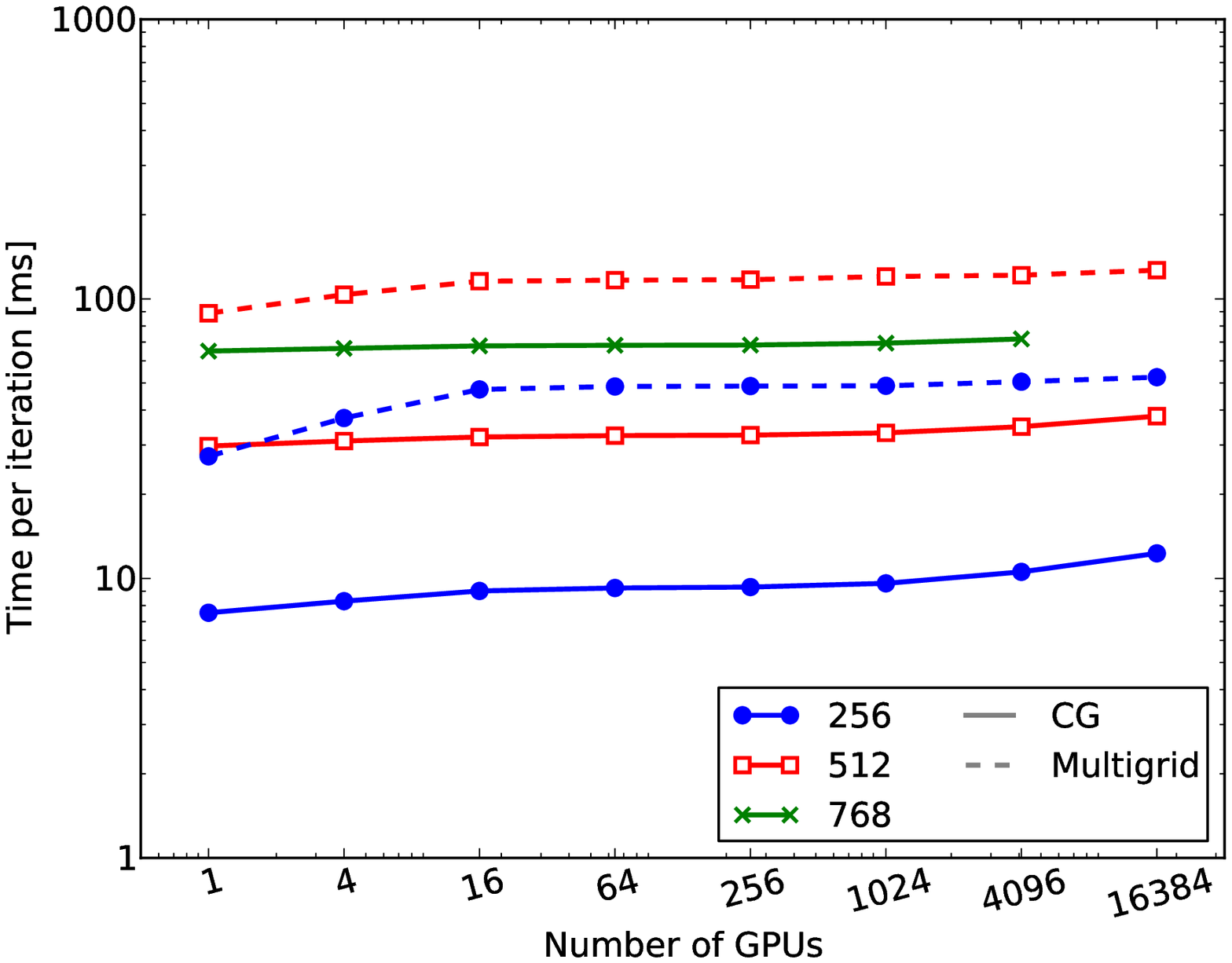}
  \includegraphics[width=0.45\linewidth]{\figdir/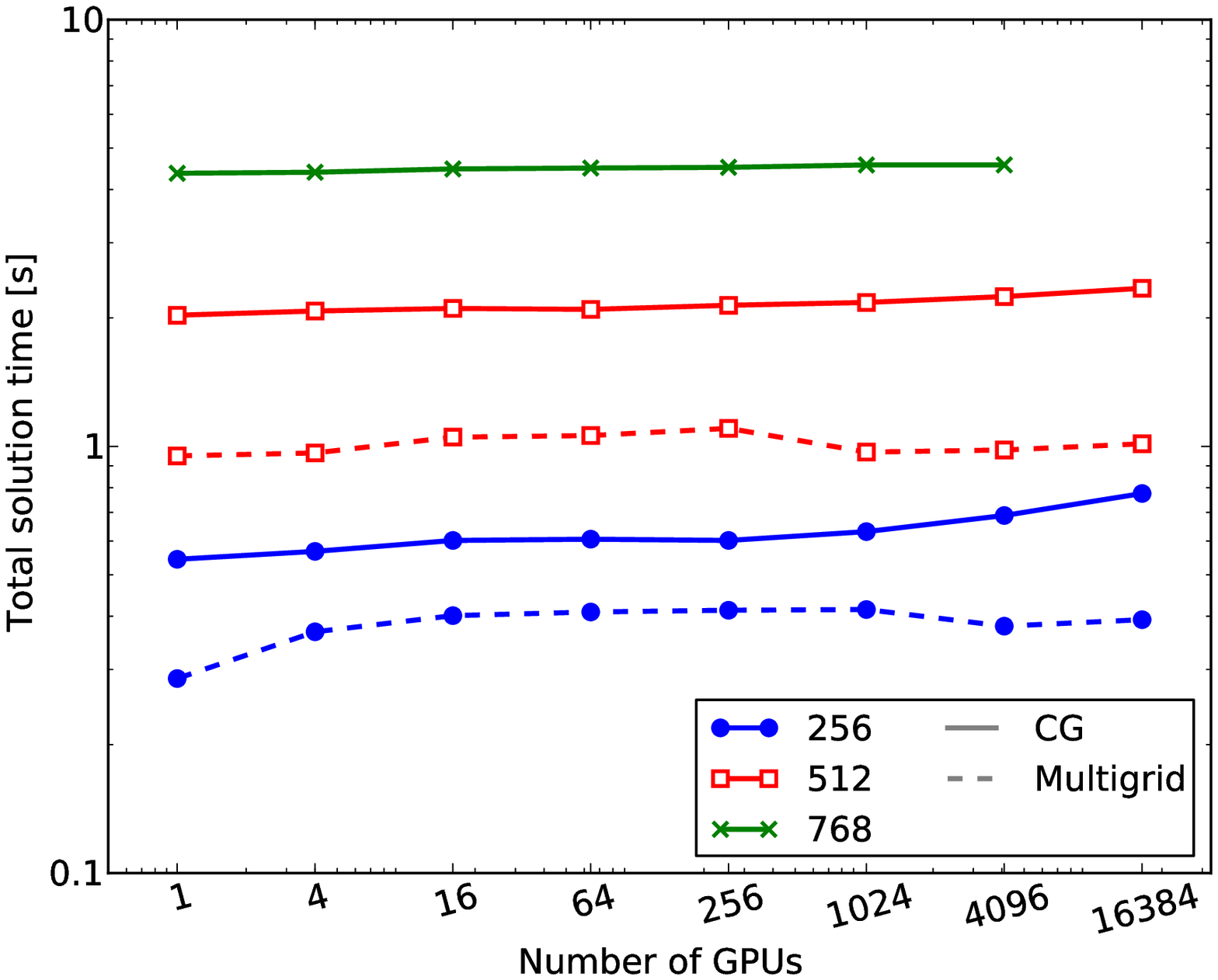}
  \caption{Weak scaling of the time per iteration (left) and total solution time (right) on different numbers of GPUs on Titan.}
  \label{fig:WeakScalingTime}
 \end{center}
\end{figure}

We finally carried out a series of weak scaling runs on Titan. For each of these runs the local problem size was kept fixed at $n_x\times n_y \times n_z$, such that the global problem size grows linearly with the number of GPUs. The resulting numbers of cores and global problem sizes are shown in Table \ref{tab:Problemsizes}. The largest problem solved has half a trillion ($0.55\cdot 10^{12}$) unknowns.
\begin{table}
  \begin{center}
    \renewcommand{\arraystretch}{1.25}
\begin{tabular}{|r|rrr|rrr|}
\hline
 & \multicolumn{2}{c}{\#GPU cores} & \# CPU cores & \multicolumn{3}{|c|}{Local problem size $n_x$}\\
\# sockets ($p$) & Titan & EMERALD & HECToR & 256 & 512 & 768\\\hline\hline
1 & 2,688 & 512 & 16 & $8.39\cdot 10^{6}$ & $3.36\cdot 10^{7}$ & $7.55\cdot 10^{7}$\\
4 & 10,752 & 2,048 & 64 & $3.36\cdot 10^{7}$ & $1.34\cdot 10^{8}$ & $3.02\cdot 10^{8}$\\
16 & 43,008 & 8,192 & 256 & $1.34\cdot 10^{8}$ & $5.37\cdot 10^{8}$ & $1.21\cdot 10^{9}$\\
64 & 172,032 & 32,768 & 1,024 & $5.37\cdot 10^{8}$ & $2.15\cdot 10^{9}$ & $4.83\cdot 10^{9}$\\
256 & 688,128 & --- & 4,096 & $2.15\cdot 10^{9}$ & $8.59\cdot 10^{9}$ & $1.93\cdot 10^{10}$\\
1024 & 2,752,512 & --- & 16,384 & $8.59\cdot 10^{9}$ & $3.44\cdot 10^{10}$ & $7.73\cdot 10^{10}$\\
4096 & 11,010,048 & --- & 65,536 & $3.44\cdot 10^{10}$ & $1.37\cdot 10^{11}$ & $3.09\cdot 10^{11}$\\
16384 & 44,040,192 & --- & --- & $1.37\cdot 10^{11}$ & $5.50\cdot 10^{11}$ & ---\\
\hline
\end{tabular}
    \caption{Global number of degrees of freedom $p\times n_x\times n_y\times n_z$ for different numbers of processors $p$ and local problem sizes $n_x=n_y$. The number of vertical columns is always $n_z=128$.}
    \label{tab:Problemsizes}
  \end{center}
\end{table}
The weak scaling of the time per iteration and the total solution time is plotted in Figure \ref{fig:WeakScalingTime}. After a slight initial increase all GPU solvers scale very well up to 16384 GPUs and, as expected, the scalability increases with the problem size due to the resulting better calculation / communication ratio. 
\begin{figure}
 \begin{center}
  \includegraphics[width=0.45\linewidth]{\figdir/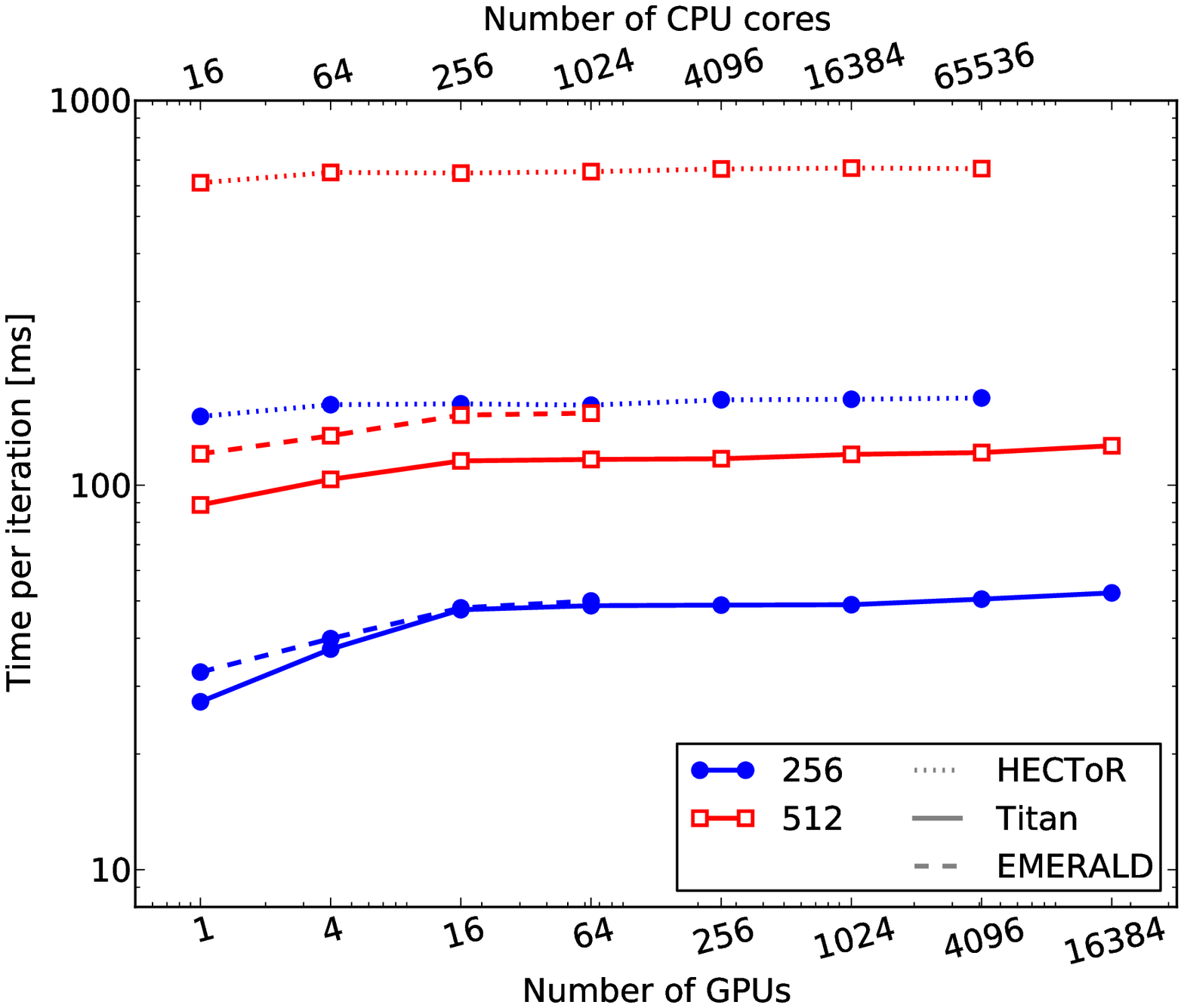}
  \includegraphics[width=0.45\linewidth]{\figdir/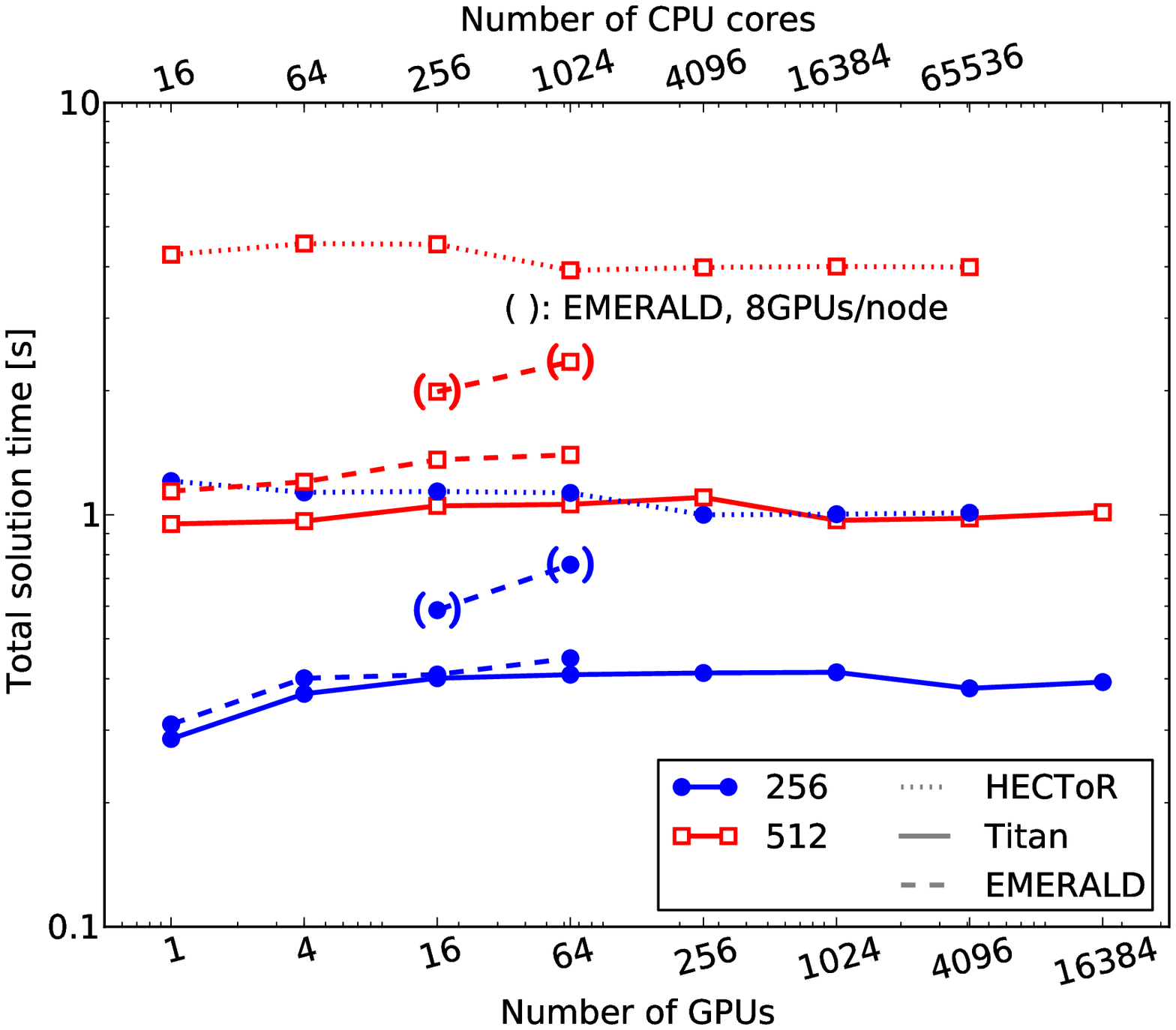}
  \caption{Time per iteration (left) and total solution time (right) of the multigrid solver on different clusters. Results obtained on EMERALD with 8 GPUs per node are shown in brackets.}
  \label{fig:WeakScalingComparison}
 \end{center}
\end{figure}
In Figure \ref{fig:WeakScalingComparison} we compare the performance of the multigrid solver for two problem sizes on all three computer systems. On the EMERALD cluster two memory links have to be shared between all GPUs on a node and we found that we can achieve the best performance if we only use two GPUs per node, all results shown in Figure \ref{fig:WeakScalingComparison}, except for the ones which are specifically marked, are obtained with this configuration. In summary the GPU implementation on Titan is roughly a factor 4 faster than the CPU implementation on HECToR (comparing one K20X GPU card to one 16 core AMD CPU) or, put differently, it is possible to solve a four times larger problem in the same total runtime. For the $512$ problem we find that the CG solver on the GPU is about 6 times faster than the CPU implementation (not shown here).

\begin{figure}
 \begin{center}
  \includegraphics[width=0.45\linewidth]{\figdir/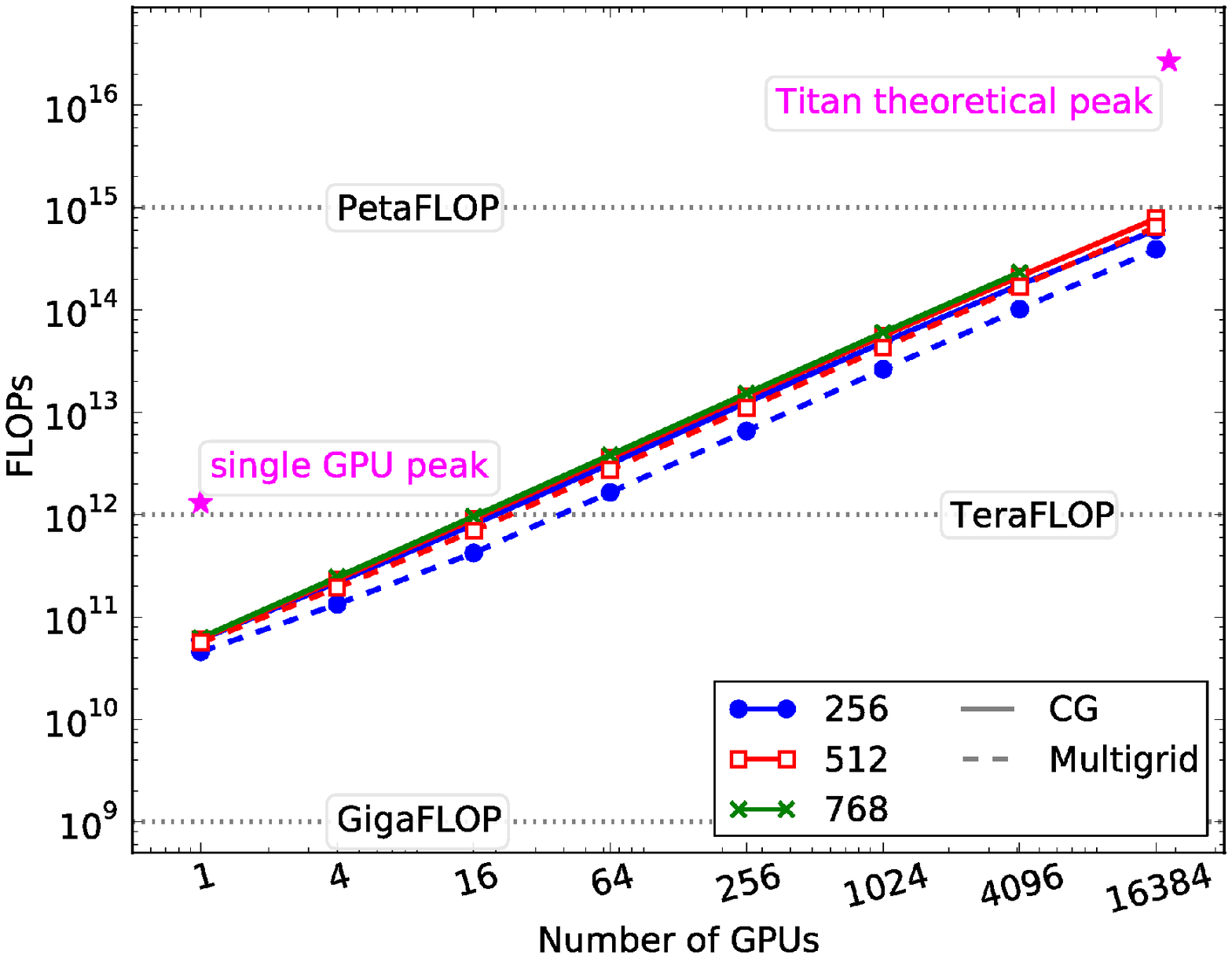}
  \includegraphics[width=0.45\linewidth]{\figdir/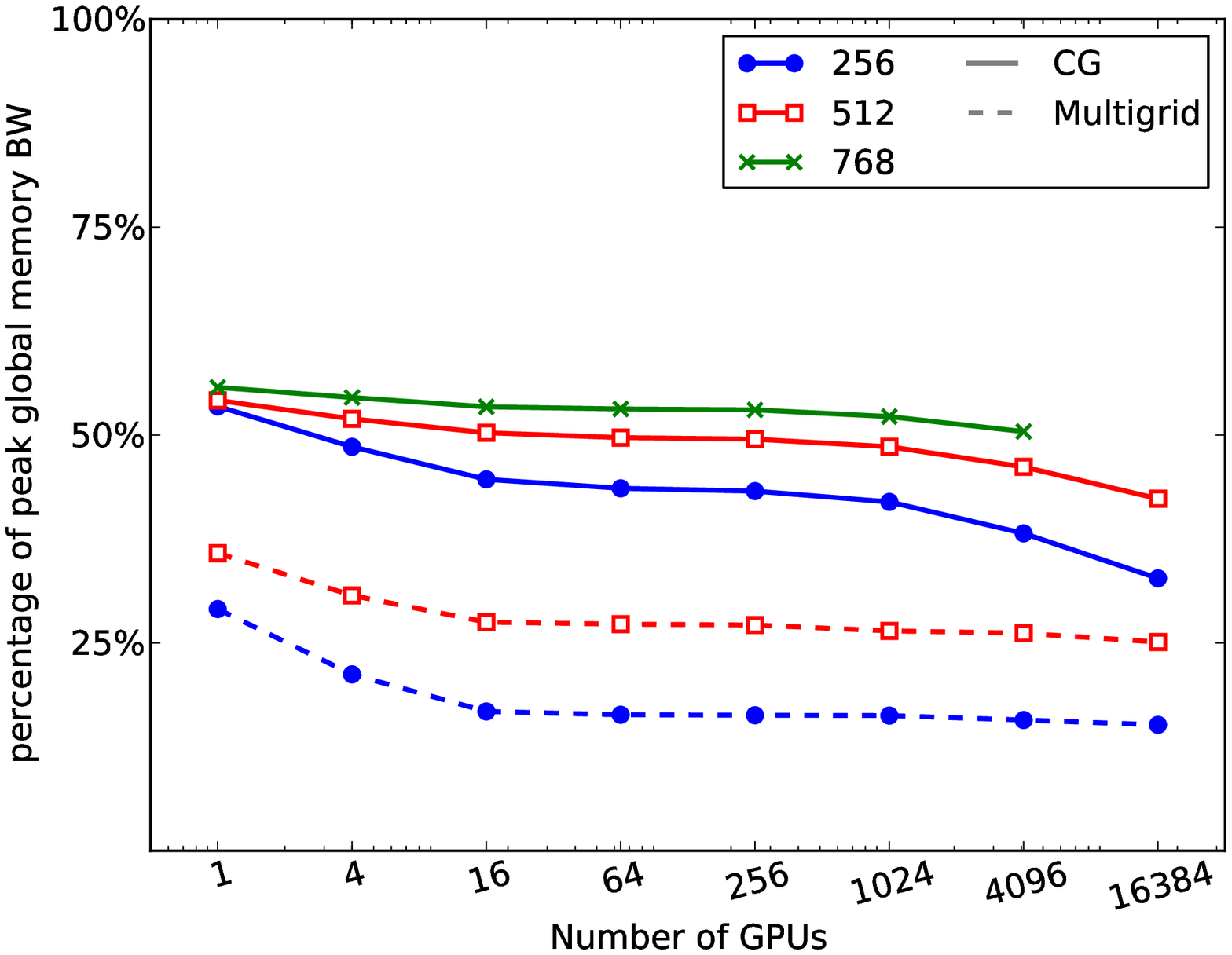}
  \caption{Floating point performance and percentage of utilised peak global memory bandwidth for different numbers of GPUs and local problem sizes $n_x$ on Titan.}
  \label{fig:FLOPandMEMPerformance}
 \end{center}
\end{figure}

The absolute performance on Titan both in terms of the global memory bandwidth and in terms of floating point operations per second was also quantified as described in Section \ref{sec:SingleGPUPerformance}. The absolute floating point performance is plotted in Figure \ref{fig:FLOPandMEMPerformance} (left) for different problem sizes and numbers of GPUs. On one GPU the CG solver on the $512$ problem can utilise about 5 \% of the peak performance and on the largest core count it can still use 3 \% of the peak FLOP rate of the entire machine.
As both algorithms are memory bound, a more meaningful measure for the performance of the solvers is the global memory bandwidth. According to Figure \ref{fig:FLOPandMEMPerformance} the CG solver can utilise between $30\%$ and $60\%$ of the theoretical peak global memory bandwidth. For the multigrid solver this number is smaller and in the range $15\%$ - $25\%$ on 16384 GPUs and $25\%$ - $35\%$ on one GPU. We stress again that we measure the ``useful bandwidth'' and obtaining a value close to $100\%$ is very hard to achieve in practice.

It is also interesting to compare the performance on the two GPU systems. Although the difference in floating point performance between the cards is a factor two, one multigrid iteration on the Fermi card is only $20\%-35\%$ slower than on the Kepler GPU, which is closer to what is expected due to the ratio between the global memory bandwidths on the two different cards ($\BWmem(K20X)/\BWmem(M2090)=(250\GBytes)/(177\GBytes)\approx 1.4$). For the same number of total GPUs, using all 8 GPUs on an EMERALD node (instead of only using 2 per node) leads to a significant increase in the total runtime as all cards have to share the host-device bandwidth (see the additional data in Figure \ref{fig:WeakScalingComparison}). This is expected as for the multigrid solver the cost of copying the right hand side to the GPU and copying the final solution back to the CPU is not negligible (see Table \ref{tab:SingleGPUTiming}).
\begin{table}
  \begin{center}
  \begin{tabular}{|r|c|rr|rr|}
    \hline
    \# GPUs & \# columns per GPU & \multicolumn{2}{|c|}{CG} & \multicolumn{2}{|c|}{Multigrid}\\
    & & $t_{\operatorname{solve}} [s]$ & parallel efficiency & $t_{\operatorname{solve}} [s]$ & parallel efficiency\\
    \hline\hline
    256  & $512\times512=262,144$ & 2.141 & ---    & 1.102 & ---    \\
    1024 & $256\times256=65,536$ & 0.631 & 84.8\% & 0.415 & 66.4\% \\
    4096 & $128\times128=16,384$ & 0.541 & 24.8\% & 0.250 & 27.6\% \\
    \hline
  \end{tabular}
  \caption{Strong scaling of the total runtime for a problem of size $8192\times 8192 \times 128$ on Titan. The parallel efficiency is given relative to the 256 GPU run.}
  \label{tab:StrongScaling}
  \end{center}
\end{table}

While we have not attempted to carry out detailled strong scaling runs, we report in Tab. \ref{tab:StrongScaling} some preliminary results for a problem of size $8192\times 8192\times128=8.6\cdot 10^{9}$ unknowns on $256$, $1024$ and $4096$ GPUs; the results were obtained on Titan. The efficiency is still around $85\%$ for CG and $66\%$ for the multigrid solver when the number of GPUs is increased from 256 to 1024, i.e. when the number of local columns on one GPU has gone from 512 x 512 to 256 x 256. When the number of GPUs is increased further to 4096, the efficiency (measured relative to the 256 GPU run) drops to around $25\%$ for both algorithms. On 4096 GPUs the number of threads per GPU is $128\times 128=16384$. This has to be compared to the number of CUDA cores on the Kepler card which is 2688, i.e. each core will process only around 6 columns. As the local problem size is reduced further the number of columns per core will drop below 1 and it becomes impossible to exploit the full capacity of the GPU with our current implementation.
\section{Conclusion and outlook}\label{sec:Conclusion}
In this article we described massively parallel and efficient CUDA-C implementations of two memory bound iterative solvers for anisotropic PDEs. Equations of this type are encountered in many areas of geophysical modelling and we focus on the pressure correction equation which arises from semi-implicit semi-Lagrangian time stepping in atmospheric forecast models. The biggest gains can be achieved by choosing the algorithmically most efficient solver tailored to the problem, which in this case is a tensor-product geometric multigrid solver. It is about twice as fast as a preconditioned Conjugate Gradient method. We demonstrated the excellent absolute performance (measured in terms of the achieved memory bandwidth) of our solvers on Kepler GK110 GPUs and showed the very good weak scaling on up to 16384 GPUs of the Titan supercomputer. The GPU implementation is about a factor four faster than the CPU code on HECToR. 

Although we believe that our implementation is close to optimal there are still potential improvements that could be considered. In this article we only give preliminary results for strong scaling of our solvers. In the strong scaling limit the number of threads will eventually be too small to utilise the full computational power of the GPU. This could be improved by exposing more parallelism in the algorithm, for example by using a parallel tridiagonal solver as described below. Furthermore the overhead from parallel communcations between GPUs can be ``hidden'' by overlapping the halo exchange with computations. For this the domain is split up into an interior part and a boundary. After the calculations have been completed on the boundary, an asynchronous halo exchange is posted and calculations are continued for the interior part of the domain. While on a CPU this approach is straighforward, on a GPU the calculations on the boundary degrees of freedom will only be efficient if it is large enough in the $x$-direction to allow contiguous memory access in this direction. This is a problem in particular for the multigrid method on the coarse levels which are already very small. In some preliminary experiments we were not able to achieve any speedups with this technique.

Part of the reason for the poorer computational efficiency of the multigrid method is that the grids on the coarse levels are so small that the GPU can not be fully utilised, as we only parallelise in the horizontal direction due to the dependency in the tridiagonal solver. The amount of parallelism can be increased if we can assign several threads to work on each vertical column, by using a parallel tridiagonal solver such as cyclic reduction or a substructuring algorithm (see also \cite{Laszlo2014}). Preliminary experiments with the subtructuring method have shown that some gains with speedups of up to a factor two can be achieved on the coarser multigrid levels.

Currently our entire solver is implemented on the GPU, which means that the
host CPU is idle. This is wasteful and could potentially be improved by
splitting the work between the two processors. The obvious way of doing this
is via splitting the domain between the CPU and GPU, as described for a
shallow water solver in \cite{Yang2013}. For the multigrid solver an
alternative approach might be to process the finer levels on the GPU and the
coarse levels on the CPU, thus minimising the amount of data that is copied between host and device. This could for example be done by using an additive multigrid algorithm, and potential benefits have to be balanced against worse algorithmic performance of the method.
\section*{Acknowledgements}
We are grateful to Benson Muite for his help with porting and running the code on the Titan supercomputer and for kindly making part of his compute time allocation on the machine available for this project. We would like to thank Mauro Bianco (CSCS, Switzerland) for his help with using the GCL library and Mike Giles and Istv\'{a}n Reguly (Oxford) for useful discussions.

This work was funded as part of the NERC project on ``Next Generation Weather
and Climate Prediction'' (NGWCP), grant number NE/K006762/1 and was supported
also by European Regional Development Fund through the Estonian Centre of
Excellence in Computer Science and the Estonian Science Foundation grant 9019.
We used resources of the Oak Ridge Leadership Computing Facility at the Oak Ridge National Laboratory, which is supported by the Office of Science of the U.S. Department of Energy under Contract No. DE-AC05-00OR22725; the runs on Titan were carried out under DD Project CSC113. 
In addition we made use of the facilities of HECToR, the UK's national high-performance computing service, which is provided by UoE HPCx Ltd at the University of Edinburgh, Cray Inc and NAG Ltd, and funded by the Office of Science and Technology through EPSRC's High End Computing Programme. We would also like to acknowledge use of the EMERALD High Performance Computing facility provided via the Centre for Innovation (CfI). The CfI is formed from the universities of Bristol, Oxford, Southampton and UCL in partnership with STFC Rutherford Appleton Laboratory.

\bibliographystyle{unsrt}

\begin{thebibliography}{10}

\bibitem{Mueller2013b}
Eike M\"{u}ller and Robert Scheichl.
\newblock {Massively parallel solvers for elliptic PDEs in numerical weather-
  and climate prediction}.
\newblock {\em accepted for publication in Q. J. Roy. Meteor. Soc.}, 2014.

\bibitem{BoermHiptmair1999}
S.~B\"{o}rm and R.~Hiptmair.
\newblock Analysis of tensor product multigrid.
\newblock {\em Numer. Algorithms}, 26:200--1, 1999.

\bibitem{Blatt07}
Markus Blatt and Peter Bastian.
\newblock {The Iterative Solver Template Library}.
\newblock In {\em {Lecture Notes in Computer Science}}, volume 4699, pages
  666--675. Springer-Verlag, Berlin, Heidelberg, 2007.

\bibitem{Blatt10}
Markus Blatt.
\newblock {A Parallel Algebraic Multigrid Method for Elliptic Problems with
  Highly Discontinuous Coefficients (PhD thesis)}.
\newblock {\em Heidelberg}, 2010.

\bibitem{Henson2000}
Van~Emden Henson and Ulrike~Meier Yang.
\newblock {BoomerAMG: a Parallel Algebraic Multigrid Solver and
  Preconditioner}.
\newblock {\em Applied Numerical Mathematics}, 41:155--177, 2000.

\bibitem{top500.org}
Hans Meuer, Erich Strohmaier, Jack Dongarra, and Horst Simon.
\newblock {Top500 list of supercomputers, November 2013}.
\newblock \url{http://www.top500.org/lists/2014/06/}.
\newblock Accessed: 13 Aug 2014.

\bibitem{TeslaDatasheet}
nVidia Corporation.
\newblock {nVidia Tesla GPU Accelerators data sheet}.
\newblock \url{http://www.nvidia.com/object/tesla-servers.html}.
\newblock Accessed: 8 Jan 2014.

\bibitem{Skamarock1997}
W~C Skamarock, P~K Smolarkiewicz, and J~B Klemp.
\newblock {Preconditioned conjugate-residual solvers for Helmholtz equations in
  nonhydrostatic models}.
\newblock {\em Mon. Weather Rev.}, 125(4):587--599, 1997.

\bibitem{Thomas1997}
S.J. Thomas, A.V. Malevsky, M.~Desgagné, R.~Benoit, P.~Pellerin, and M.~Valin.
\newblock Massively parallel implementation of the mesoscale compressible
  community model.
\newblock {\em Parallel Comput.}, 23:2143 -- 2160, 1997.

\bibitem{Qaddouri2003}
Abdessamad Qaddouri and Jean C\^{o}t\'{e}.
\newblock {Preconditioning for an Iterative Elliptic Solver on a Vector
  Processor}.
\newblock In José Palma, A.~Sousa, Jack Dongarra, and Vicente Hernández,
  editors, {\em High Performance Computing for Computational Science VECPAR
  2002}, volume 2565 of {\em Lecture Notes in Computer Science}, pages
  451--455. Springer, Berlin, 2003.

\bibitem{Davies05}
T.~Davies, M.~J.~P. Cullen, A.~J. Malcolm, M.~H. Mawson, A.~Staniforth, A.~A.
  White, and N.~Wood.
\newblock A new dynamical core for the {Met Office's} global and regional
  modelling of the atmosphere.
\newblock {\em Q. J. Roy. Meteor. Soc.}, 131(608):1759--1782, 2005.

\bibitem{Press2007}
William~H. Press, Saul~A. Teukolsky, William~T. Vetterling, and Brian~P.
  Flannery.
\newblock {\em {Numerical Recipes: The Art of Scientific Computing, 3rd
  Edition}}.
\newblock Cambridge University Press, New York, 2007.

\bibitem{Mueller2013a}
Eike M\"{u}ller, Xu~Guo, Robert Scheichl, and Sinan Shi.
\newblock {Matrix-free GPU implementation of a preconditioned Conjugate
  Gradient solver for anisotropic elliptic PDEs}.
\newblock {\em to appear in Computation and Visualization in Science}, Feb
  2014.

\bibitem{bianco2013interface}
Mauro Bianco.
\newblock An interface for halo exchange pattern.
\newblock \url{http://www.prace-ri.eu/IMG/pdf/wp86.pdf}, 2013.
\newblock Accessed: 11 Jan 2014.

\bibitem{Dedner2014}
Andreas Dedner, Eike Mueller, and Robert Scheichl.
\newblock {Efficient multigrid preconditioners for anisotropic elliptic PDEs in
  geophysical modelling}.
\newblock submitted to SIAM Journal on Scientific Computing (SISC), 2014.

\bibitem{Bolz03}
Jeff Bolz, Ian Farmer, Eitan Grinspun, and Peter Schr\"{o}der.
\newblock {Sparse Matrix Solvers on the GPU: Conjugate Gradients and
  Multigrid}.
\newblock {\em ACM Transactions on Graphics}, 22:917--924, 2003.

\bibitem{Goodnight2005}
Nolan Goodnight, Cliff Woolley, Gregory Lewin, David Luebke, and Greg
  Humphreys.
\newblock {A multigrid solver for boundary value problems using programmable
  graphics hardware}.
\newblock In {\em ACM SIGGRAPH 2005 Courses}, SIGGRAPH '05, New York, NY, USA,
  2005. ACM.

\bibitem{Menon2007}
S~Menon and JB~Perot.
\newblock {Implementation of an efficient conjugate gradient algorithm for
  Poisson solutions on graphics processors}.
\newblock In {\em Proceedings of the 2007 Meeting of the Canadian CFD Society,
  Toronto Canada}, 2007.

\bibitem{cevahir2009fast}
Ali Cevahir, Akira Nukada, and Satoshi Matsuoka.
\newblock {Fast conjugate gradients with multiple GPUs}.
\newblock In {\em Computational Science--ICCS 2009}, pages 893--903. Springer
  Verlag, Berlin, Heidelberg, 2009.

\bibitem{Knittel2010}
M.~Ament, G.~Knittel, D.~Weiskopf, and W.~Strasser.
\newblock {A Parallel Preconditioned Conjugate Gradient Solver for the Poisson
  Problem on a Multi-GPU Platform}.
\newblock In {\em {Parallel, Distributed and Network-Based Processing (PDP),
  2010, 18th Euromicro International Conference on}}, pages 583 --592, feb.
  2010.

\bibitem{cevahir2010high}
Ali Cevahir, Akira Nukada, and Satoshi Matsuoka.
\newblock {High performance conjugate gradient solver on multi-GPU clusters
  using hypergraph partitioning}.
\newblock {\em Computer Science-Research and Development}, 25(1-2):83--91,
  2010.

\bibitem{Georgescu2010}
Serban Georgescu and Hiroshi Okuda.
\newblock {Conjugate gradients on multiple GPUs}.
\newblock {\em International Journal for Numerical Methods in Fluids},
  64(10-12):1254--1273, 2010.

\bibitem{griebel2010multi}
Michael Griebel and Peter Zaspel.
\newblock {A multi-GPU accelerated solver for the three-dimensional two-phase
  incompressible Navier-Stokes equations}.
\newblock {\em Computer Science-Research and Development}, 25(1-2):65--73,
  2010.

\bibitem{jacobsen2010mpi}
Dana~A Jacobsen, Julien~C Thibault, and Inanc Senocak.
\newblock {An MPI-CUDA implementation for massively parallel incompressible
  flow computations on multi-GPU clusters}.
\newblock In {\em 48th AIAA Aerospace Sciences Meeting and Exhibit}, volume~16,
  2010.

\bibitem{feng2010parallel}
Zhuo Feng and Zhiyu Zeng.
\newblock {Parallel multigrid preconditioning on graphics processing units
  (GPUs) for robust power grid analysis}.
\newblock In {\em Proceedings of the 47th Design Automation Conference}, pages
  661--666. ACM, 2010.

\bibitem{Geveler2011}
Markus Geveler, Dirk Ribbrock, Dominik G{\"o}ddeke, Peter Zajac, and Stefan
  Turek.
\newblock {\em {Efficient finite element geometric multigrid solvers for
  unstructured grids on GPUs}}.
\newblock Techn. Univ., Fak. f{\"u}r Mathematik, 2011.

\bibitem{jacobsen2011full}
Dana~A Jacobsen and Inanc Senocak.
\newblock {A full-depth amalgamated parallel 3D geometric multigrid solver for
  GPU clusters}.
\newblock In {\em 49th AIAA Aerospace Science Meeting}, 2011.

\bibitem{Bell2008}
Nathan Bell and Michael Garland.
\newblock Efficient sparse matrix-vector multiplication on {CUDA}.
\newblock NVIDIA Technical Report NVR-2008-004, NVIDIA Corporation, December
  2008.

\bibitem{Davis2011}
Timothy~A. Davis and Yifan Hu.
\newblock {The University of Florida Sparse Matrix Collection}.
\newblock {\em ACM Trans. Math. Softw.}, 38(1):1:1--1:25, December 2011.

\bibitem{haase2010parallel}
Gundolf Haase, Manfred Liebmann, Craig~C Douglas, and Gernot Plank.
\newblock {A parallel algebraic multigrid solver on graphics processing units}.
\newblock In {\em High performance computing and applications}, pages 38--47.
  Springer, 2010.

\bibitem{Brannick2013}
James Brannick, Yao Chen, Xiaozhe Hu, and Ludmil Zikatanov.
\newblock {Parallel Unsmoothed Aggregation Algebraic Multigrid Algorithms on
  GPUs}.
\newblock In Oleg~P. Iliev, Svetozar~D. Margenov, Peter~D Minev, Panayot~S.
  Vassilevski, and Ludmil~T Zikatanov, editors, {\em Numerical Solution of
  Partial Differential Equations: Theory, Algorithms, and Their Applications},
  volume~45 of {\em Springer Proceedings in Mathematics \& Statistics}, pages
  81--102. Springer New York, 2013.

\bibitem{Goeddeke2008}
Dominik G{\"o}ddeke, Robert Strzodka, Jamaludin Mohd-Yusof, Patrick McCormick,
  Hilmar Wobker, Christian Becker, and Stefan Turek.
\newblock Using {GPUs} to improve multigrid solver performance on a cluster.
\newblock {\em International Journal of Computational Science and Engineering
  (IJCSE)}, 4(1):36--55, 2008.

\bibitem{Yang2013}
Chao Yang, Wei Xue, Haohuan Fu, Lin Gan, Linfeng Li, Yangtong Xu, Yutong Lu,
  Jiachang Sun, Guangwen Yang, and Weimin Zheng.
\newblock {A Peta-scalable CPU-GPU Algorithm for Global Atmospheric
  Simulations}.
\newblock {\em SIGPLAN Not.}, 48(8):1--12, February 2013.

\bibitem{Gmeiner2014}
Bj\"{o}rn Gmeiner, Harald K\"{o}stler, Markus St\"{u}rmer, and Ulrich R\"{u}de.
\newblock {Parallel multigrid on hierarchical hybrid grids: a performance study
  on current high performance computing clusters}.
\newblock {\em Concurrency and Computation: Practice and Experience},
  26(1):217--240, 2014.

\bibitem{Trottenberg2001}
U~Trottenberg, C~W Oosterlee, and A~Sch\"uller.
\newblock {\em {Multigrid}}.
\newblock Academic Press, San Diego, London, Sydney, Tokyo, 2001.

\bibitem{Wood2013}
N.~Wood, A.~Staniforth, A.~White, T.~Allen, M.~Diamantakis, M.~Gross,
  T.~Melvin, C.~Smith, S.~Vosper, M.~Zerroukat, and J.~Thuburn.
\newblock An inherently mass-conserving semi-implicit semi-{L}agrangian
  discretisation of the deep-atmosphere global nonhydrostatic equations.
\newblock {\em Q. J. Roy. Meteor. Soc.}, 2013.
\newblock Published online December 4th 2013.

\bibitem{Sadourny1972}
Robert Sadourny.
\newblock {Conservative Finite-Difference Approximations of the Primitive
  Equations on Quasi-Uniform Spherical Grids}.
\newblock {\em {Mon. Weather Rev.}}, 100(2):136--144, 1972.

\bibitem{Saad2003}
Yousef Saad.
\newblock {\em {Iterative Methods for Sparse Linear Systems, Second Edition}}.
\newblock Society for Industrial and Applied Mathematics, Philadelphia, 2003.

\bibitem{Briggs2000}
W.~L. Briggs, V.~E. Henson, and S.~F. McCormick.
\newblock {\em {A Multigrid Tutorial}}.
\newblock Society for Industrial and Applied Mathematics, Philadelphia, 2000.

\bibitem{OpteronDatasheet}
AMD Corporation.
\newblock {}.
\newblock
  \url{http://www.amd.com/uk/products/server/processors/6000-series-platform/6200/Pages/6200-series-processors.aspx}.
\newblock Accessed: 12 Jan 2014.

\bibitem{Harris2013}
Mark Harris.
\newblock {An Efficient Matrix Transpose in CUDA C/C++}.
\newblock
  \url{http://devblogs.nvidia.com/parallelforall/efficient-matrix-transpose-cuda-cc/}.
\newblock Accessed: 7 Jan 2014.

\bibitem{Laszlo2014}
Endre Laszlo.
\newblock {Efficient Solution of Multiple Scalar and Block-Tridiagonal
  Equations}.
\newblock GPU Technology Conference, 2014.

\end{thebibliography}

\end{document}